\documentclass[aps,pra,reprint,superscriptaddress,nofootinbib]{revtex4-2}

\usepackage[T1]{fontenc}
\usepackage[utf8]{inputenc}
\usepackage{microtype}
\usepackage{amsthm}
\usepackage{ragged2e}
\usepackage{bm}
\usepackage{afterpage}
\usepackage{placeins}

\usepackage{amsmath, amssymb, amsfonts, mathtools}
\usepackage{physics}
\usepackage{siunitx}
\usepackage{empheq}
\usepackage{cases}

\usepackage{graphicx}
\usepackage{subcaption}
\usepackage{float}
\usepackage{booktabs}
\usepackage{multirow}
\usepackage{array}
\usepackage{tabularx}
\usepackage{threeparttable}
\usepackage{caption}
\usepackage{adjustbox}

\usepackage{algorithm}
\usepackage{algpseudocode}
\usepackage{listings}
\usepackage{xcolor}
\usepackage{tikz}

\usepackage[hidelinks]{hyperref}
\usepackage[nameinlink,noabbrev,capitalise]{cleveref}

\usepackage{enumitem}
\usepackage{url}
\usepackage{nicefrac}
\usepackage{orcidlink}
\definecolor{codegray}{gray}{0.95}
\definecolor{codecomment}{RGB}{0,100,0}
\definecolor{codekeyword}{RGB}{0,0,150}

\renewcommand{\dd}{\mathrm{d}}

\makeatletter
\newenvironment{widetableblock}{%
  \par\ignorespaces
  \onecolumngrid
  \vskip6\p@
  \prep@math@patch
}{%
  \par
  \vskip6\p@
  \twocolumngrid\global\@ignoretrue
  \@endpetrue
}
\makeatother

\newtheorem{proposition}{Proposition}

\theoremstyle{definition}

\renewcommand{\dd}{\mathrm{d}}


\begin{document}

\title{
Efficient Quantum Simulation of Variable-Coefficient Transport with Continuous Source Injection
}

\author{Mohammad Mehedi Hasan Akash\,\orcidlink{0000-0001-7520-0531}}
\author{Turag Dev}
\affiliation{Department of Mechanical and Aerospace Engineering, Joint College of Engineering, FAMU-Florida State University, Tallahassee, FL 32310}
\author{Nhat-Quang Nguyen\,\orcidlink{https://orcid.org/0009-0003-2981-6679}}
\affiliation{Department of Physics, Florida State University, Tallahassee, FL 32306}
\affiliation{FSU Quantum Initiative, Florida State University, Tallahassee, FL 32306}

\author{Yanzhu Chen\,\orcidlink{https://orcid.org/0000-0001-5589-9197}}
\affiliation{Department of Physics, Florida State University, Tallahassee, FL 32306}
\affiliation{FSU Quantum Initiative, Florida State University, Tallahassee, FL 32306}

\author{Huixuan Wu\,\orcidlink{https://orcid.org/0000-0003-1999-7114}}
\affiliation{Department of Mechanical and Aerospace Engineering, Joint College of Engineering, FAMU-Florida State University, Tallahassee, FL 32310}

\author{Kourosh Shoele\,\orcidlink{https://orcid.org/0000-0003-2810-0065}}
\email[Corresponding author: ]{kshoele@eng.famu.fsu.edu}
\affiliation{Department of Mechanical and Aerospace Engineering, Joint College of Engineering, FAMU-Florida State University, Tallahassee, FL 32310}

\date{\today}

\begin{abstract}

Quantum time-marching algorithms for transport partial differential equations commonly represent variable coefficients and forcing using register-expanding dilations, block-encoding oracles, or repeated postselection. We present an alternative algorithm for a forced variable-coefficient advection-diffusion equation in flow-inspired skew-symmetric form, treated as a prototype for flow-momentum equations, that incorporates spatially varying velocity, viscous dissipation, and persistent source injection with a peak logical requirement of $n_q+1$ qubits, where $n_q$ is the number of qubits encoding the spatial field. A centered skew-symmetric discretization renders the advection operator strictly skew-Hermitian for arbitrary velocity profiles, enabling an ancilla-free unitary implementation using a Gray-code Trotter sequence of controlled-$R_y$ rotations. Diffusion is applied in the Fourier basis through a uniformly controlled rotation on one postselected ancilla, which is measured, reset, and reused between the two diffusion half-steps, while the source is incorporated classically within second-order Strang splitting. Statevector simulations for $N=16$ and $32$ recover second-order temporal convergence relative to high-accuracy classical solutions, while Richardson extrapolation delivers fourth-order accuracy and reduces the required kernel calls by factors of four to fourteen. Independent tests through $N=256$ confirm second-order spatial consistency. We further show analytically and numerically that the per-step ancilla failure probability is directly proportional to the instantaneous viscous dissipation rate, making the postselection cost self-regulating over a fifty-fold viscosity range. Stable evolution is demonstrated for $5\times10^4$ time steps without observable secular error growth, while Gray-code advection accounts for $71$--$95\%$ of the transpiled controlled-NOT gates. The resulting fixed-width kernel provides a qubit-efficient building block for near-term hardware studies, although classical readout and state re-preparation remain the principal obstacles to coherent multistep evolution.

\end{abstract}

\maketitle

\section{Introduction}
Computational fluid dynamics rests on the numerical solution of partial differential equations whose discretized state vectors grow rapidly with spatial resolution, placing large-scale flow simulations among the most demanding workloads in scientific computing~\cite{leveque2007finite}. Quantum computers offer a fundamentally different way to store such fields: amplitude encoding represents $N = 2^{n_q}$ field values in the amplitudes of only $n_q$ qubits, an exponentially compact register~\cite{nielsen2010quantum} on which the discretized evolution operators of the flow can act directly \cite{lloyd1996universal,malinverno2025review,georgescu2014quantum}. Realistic transport problems, however, rarely involve uniform coefficients; advecting velocities vary in space, viscous dissipation acts across all scales, and external forcing continuously injects momentum or scalar concentration into the flow. The central question addressed in this work is whether a quantum algorithm can correctly solve a transport equation that combines a spatially varying velocity $U(x)$, viscous diffusion, and persistent source injection within a fixed qubit budget.

Quantum algorithms for partial differential equations fall broadly into several families. Quantum linear systems methods recast the discretized equation as a linear system and solve it with the Harrow-Hassidim-Lloyd algorithm and its high-precision descendants \cite{harrow2009quantum,childs2021high,berry2017quantum,costa2022optimal}; yet, these methods are natural for steady-state problems but less suited to explicit time evolution. Variational approaches alternatively encode the solution in a parametrized circuit, as applied to the one-dimensional advection-diffusion equation by Ingelmann et al.~\cite{ingelmann2024two}, yet ansatz expressivity and barren plateaus restrict the scalability of this method.  Another class of approaches uses Schr\"{o}dinger embeddings, in which the non-Hermitian transport operator is mapped to a Hamiltonian acting on an enlarged Hilbert space. Examples include Schr\"{o}dingerisation~\cite{hu2024quantum,guseynov2026quantum}, the linear-combination-of-Hamiltonian-simulations approach of Novikau and Joseph~\cite{novikau2025quantum}, and Carleman linearization~\cite{sanavio2025explicit,demirdjian2026efficient}. These methods are broadly applicable, in which with Schr\"{o}dingerisation, we can even accommodate variable coefficients and inhomogeneous terms~\cite{guseynov2026quantum}. Yet, this generality comes at the cost of register-expanding dilations and the associated ancilla overhead.

The family that is most natural for fluid time integration, and the one this work belongs to, advances the solution by time marching: the evolution operator over a step $\Delta t$ is decomposed by operator splitting into quantum circuits applied sequentially~\cite{brearley2024quantum,over2025quantum,bharadwaj2025compact,lubasch2025quantum,gaitan2020finding}, mirroring classical time-stepping schemes.
Brearley and Laizet~\cite{brearley2024quantum} embedded the discrete advection operator into a Hamiltonian simulation, handling general stencils, boundary conditions, and velocity fields through a Hermitian dilation with per-step post-selection. In this method, the construction addresses only pure advection, with neither viscous diffusion nor a source term. Over et al.~\cite{over2025quantum} block-encode the explicit advection-diffusion time-marching operator as a linear combination of an advection-like dilation and a corrective shift operator, but the shift-operator correction encodes a spatially uniform diffusion coefficient, and the system must remain homogeneous . In later work, it was extended to non-periodic boundary conditions while retaining the homogeneous setting~\cite{bengoechea2026quantum}. Bharadwaj and Sreenivasan~\cite{bharadwaj2025compact} proposed compact linear-combination-of-unitaries algorithms with reduced qubit overhead for advection-diffusion at constant velocity, with oracle and state-preparation costs that grow once coefficients vary in space. Lubasch et al.~\cite{lubasch2025quantum} combine the quantum Fourier transform with quantum singular value transformation to obtain  simple Fourier-space circuits, an approach whose Fourier-diagonal structure requires spatially uniform coefficients. Across all of these algorithms, two restrictions persist. None provides a mechanism for persistent source injection. And where spatially varying velocity is supported, it is done by embedding the advection operator in a Hermitian dilation, so that even the unitary part of the dynamics inherits an ancilla register and a per-step post-selection burden.

Incorporating persistent source injection into any of these architectures would demand additional register-expanding embeddings that challenge near-term hardware limits~\cite{preskill2018quantum}. Among the advection--diffusion time-marching constructions considered here, an explicit treatment combining spatially varying skew-symmetric advection, viscous diffusion, and persistent source injection within the present fixed-register $n_q+1$ architecture has not been demonstrated.

Recent work further broadens the quantum-PDE landscape. Alipanah et al.~\cite{alipanah2025quantum} investigated Trotterized and variational quantum-dynamics formulations of the advection--diffusion equation, including simulator and superconducting-hardware demonstrations. Sato et al.~\cite{sato2025quantum} developed an LCHS-based framework for linear PDEs with spatially varying parameters, while Guseynov et al.~\cite{guseynov2025gate} provided explicit gate-level block encodings for Hamiltonians arising from broad classes of discretized linear PDEs. Complementary complexity analyses of multidimensional drift--diffusion equations have compared Fourier-based, Hamiltonian-simulation, quantum-linear-system, and quantum-random-walk formulations~\cite{devereux2025}. These developments demonstrate that variable-coefficient and transport PDEs can be addressed through several general-purpose quantum frameworks. The distinction pursued here is narrower: the skew-symmetric variable-coefficient advective generator is exploited directly as an ancilla-free unitary, while non-unitarity is confined to the one-ancilla diffusion primitive and persistent forcing is incorporated through a deterministic classical update.

In this work, we discuss a hybrid quantum-classical algorithm for an energy-conserving variable-coefficient advection-diffusion equation with source injection. The method uses a skew-symmetric discretization of the continuum generator $-\tfrac12[U\partial_x+\partial_x(U\,\cdot)]$ inspired by the common conservative representation of convective terms in the Navier-Stokes equations {\cite{zang1991rotation,morinishi1998fully}}.
For an arbitrary sampled velocity profile $U(x)$, the discrete generator is skew-Hermitian, and its exponential is therefore exactly unitary. Consequently, in contrast to the embedding-based treatments discussed above, the advective dynamics can be implemented without dilation, ancillary degrees of freedom, or postselection.
This unitary is implemented as a Gray-code Trotterized sequence of single-target controlled-$R_y$ gates, with Gray-code ordering that reduces each nearest-neighbor coupling to a single-qubit transition. Viscous diffusion, the genuinely non-unitary part of the dynamics, is isolated in the Fourier basis and applied by a uniformly controlled $R_y$ rotation~\cite{mottonen2004} on one post-selected ancilla qubit, damping each Fourier mode individually. Source injection is handled entirely classically via Strang operator splitting~\cite{strang1968construction}, so the source step succeeds with probability 1 and incurs no ancilla overhead. Initial conditions are loaded by a Fourier state preparation based on a binary-tree decomposition~\cite{gleinig2021efficient} that encodes the $K_n$ active modes at step $n$. The homogeneous kernel requires only $n_q+1$ qubits. We validate the model against high-accuracy classical integrators of the same semi-discrete equation and report state preparation and reconstruction separately from the kernel resources.

The remainder of the paper is organized as follows. Section~\ref{sec:methodology} introduces the governing equation and its skew-symmetric operator decomposition, then develops the quantum circuit architecture, comprising sparse Fourier state preparation, uniformly controlled $R_y$ diffusion, and Gray-code Trotterized advection, together with a commutator-based error estimate and the resulting qubit and gate budget. Section~\ref{sec:results} validates the algorithm against high-accuracy classical integrators, characterizes its temporal convergence and Richardson-extrapolated accuracy, verifies the ancilla-dissipation identity linking post-selection cost to physical viscous dissipation, and reports circuit resource projections together with long-horizon stability. Section~\ref{sec:discussion} compares this algorithm relative to existing quantum transport solvers, examines the cost structure separating quantum kernel calls from the classical state-preparation and readout boundary, and outlines the outlook toward multi-step and higher-dimensional extensions. Section~\ref{sec:conclusions} summarizes the principal findings and limitations.

\section{Methodology}
\label{sec:methodology}

\subsection{ Governing Equation and Operator Splitting}
\label{sec:governing}

We consider the forced, variable-coefficient advection--diffusion equation in skew-symmetric form on the periodic domain $x\in[0,L)$,
\begin{align}
  \frac{\partial u}{\partial t}
  ={}&-\frac{1}{2}\left[
      U(x)\frac{\partial u}{\partial x}
      +\frac{\partial [U(x)u]}{\partial x}\right]
      +\nu\,\frac{\partial^2 u}{\partial x^2}
      +f(x), \notag\\
  &x\in[0,L),\qquad t\in[0,T].
  \label{eq:pde}
\end{align}
This split form is directly relevant to conservative flow simulations, as the symmetric combination of the advective and conservative forms preserves the skew-adjoint transport structure and its associated quadratic invariant, consistent with energy-conserving formulations commonly used in CFD. Here, $u(x,t)$ denotes the evolving scalar field,
$U(x)=U_0+U_1\sin(2\pi x/L)$ is the spatially varying transport velocity, with $U_0=1.0$ and $U_1=0.3$, $\nu$ is the kinematic viscosity, and $f(x)$ is a prescribed time-independent forcing term. All quantities in Eq.~\eqref{eq:pde} are nondimensional. The choice $L=1$ fixes the length scale, with $t$, $U(x)$, $\nu$, and $f(x)$ scaled accordingly.

Periodic boundary conditions, $u(0,t)=u(L,t)$, permit a uniform discretization without ghost points or one-sided stencils. We use $N=2^{n_q}$ grid points,
$x_m=mh$, with $h=L/N$ and $m=0,\ldots,N-1$, so that the discretized state is represented by $\mathbf{u}(t)\in\mathbb{C}^N$. The semi-discrete evolution is
\begin{equation}
  \frac{\mathrm{d}\mathbf{u}}{\mathrm{d}t}
  = \underbrace{\mathbf{K}_{\mathrm{adv}}\mathbf{u}}_{\text{advection}}
  + \underbrace{\mathbf{D}_{\mathrm{diff}}\mathbf{u}}_{\text{diffusion}}
  + \mathbf{f}.
  \label{eq:semidiscrete}
\end{equation}
Let $\mathbf D_c$ denote the periodic centered-difference matrix, satisfying
$\mathbf D_c^T=-\mathbf D_c$, and let
$\mathbf U=\operatorname{diag}(U(x_0),\ldots,U(x_{N-1}))$.
The advection operator is constructed as
\begin{align}
 \mathbf K_{\mathrm{adv}}
 &=-\frac12(\mathbf U\mathbf D_c+\mathbf D_c\mathbf U)
 =\frac12(\mathbf C-\mathbf C^T),\notag\\
 \mathbf C&=-\mathbf U\mathbf D_c .
 \label{eq:Kadv}
\end{align}
This skew-symmetric construction is deliberate: at the continuum level, it corresponds to
$-Uu_x-(\partial_xU)u/2$, rather than to the unsymmetrized term $-U\,u_x$ alone. Thus, the circuit implements Eq.~\eqref{eq:pde} and, in the inviscid unforced limit, retains an energy-conserving discrete advective generator.

\begin{figure}[t]
  \centering

  \includegraphics[width=\columnwidth]{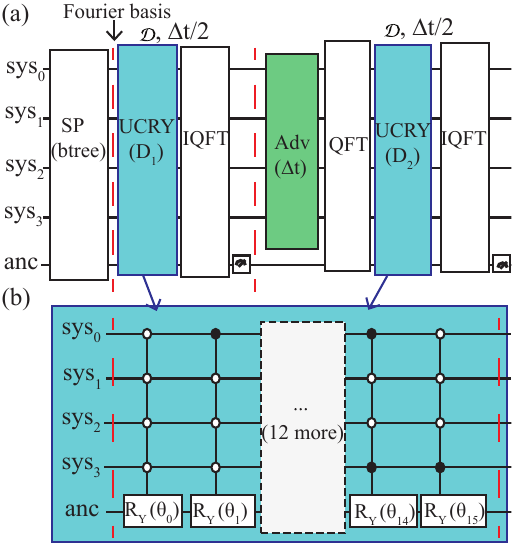}
  \vspace{2pt}

  \caption{%
    Quantum circuit for one D--A--D time step with $n_q=4$ ($N=16$ Fourier modes).
    \textbf{(a)} Circuit overview. Sparse binary-tree state preparation, SP(btree), loads the source-corrected state $|\hat{S}\rangle$ directly in the Fourier basis, eliminating a QFT before the first diffusion half-step $\mathcal{D}(\Delta t/2)$. Each diffusion block applies a uniformly controlled $R_y$ (UCRY) operation to the data register and a single ancilla, followed by an IQFT and post-selective ancilla measurement ($\otimes$). The ancilla is reset and reused for the second diffusion half-step, which is preceded by a QFT because the advection step returns the register to the position basis. The advection circuit is detailed in Fig.~\ref{fig:advection}.
    \textbf{(b)} UCRY implementation of the first diffusion half-step. For each Fourier mode $k$, the corresponding control pattern applies $R_y(\theta_k)$ to the ancilla, with
    $\theta_k=2\arccos(\lambda_k)$ and
    $\lambda_k=\exp(-\nu\kappa_k^2\Delta t/2)$.
    Filled (open) circles denote controls on $|1\rangle$ ($|0\rangle$). The rotations for $k=0,1,14,15$ are shown explicitly, with the intermediate modes represented by the dashed region. Post-selecting $|0\rangle_{anc}$ yields the diffusion-attenuated state.
    }
    \label{fig:diffusion}
\end{figure}
\begin{proposition}[Discrete advection structure]
For real grid values of $U$, $\mathbf K_{\mathrm{adv}}^\dagger =-\mathbf K_{\mathrm{adv}}$. Consequently, $\exp(t\mathbf K_{\mathrm{adv}})$ is unitary and preserves $\|\mathbf u\|_2$ for every $t$.
\end{proposition}
\begin{proof}
Because $\mathbf U^\dagger=\mathbf U$ and
$\mathbf D_c^\dagger=-\mathbf D_c$, taking the adjoint of
Eq.~\eqref{eq:Kadv} changes its sign. The exponential of
a skew-Hermitian matrix is unitary.
\end{proof}

The diffusion operator
$\mathbf D_{\mathrm{diff}} =
F^\dagger\operatorname{diag}(-\nu\kappa_k^2)F$
is diagonal in Fourier space. For the even grid sizes $N=2^{n_q}$ employed here, the signed physical wavenumbers follow the natural FFT ordering,
\begin{equation}
\kappa_k=\frac{2\pi}{L}
\begin{cases}
k, & 0\leq k < N/2,\\[2mm]
k-N, & N/2\leq k\leq N-1,
\end{cases}
\label{eq:signed_wavenumbers}
\end{equation}
where indices $k>N/2$ represent negative-frequency modes, while $k=N/2$ is the Nyquist mode with $\kappa_{N/2}=-\pi/h$. Since each Fourier mode is attenuated over a half-step by $\lambda_k=e^{-\nu\kappa_k^2\Delta t/2}\leq1$, the diffusion update is non-unitary and must be implemented using an auxiliary ancilla~\cite {harrow2009quantum} qubit.
The simulation performed here is initialized from a normalized Gaussian profile, $u(x,0) \propto \exp(-\sigma(x - x_c)^2)$, centered at $x_c = 0.5$ with width parameter $\sigma = 60$, on a domain of length $L = 1$.

At each time level, the normalized field is amplitude-encoded in the quantum register in the Fourier basis. Following the same Fourier convention used for the state-preparation target explained later in Eq.~\eqref{eq:Shat}, namely as,
\begin{equation}
  |\psi(t)\rangle = \sum_{k=0}^{N-1} \hat{\psi}_k(t)\,|k\rangle,
  \qquad
  \hat{\psi}_k(t) = \frac{\overline{\left[\mathcal{F}\{\mathbf{u}(t)/\|\mathbf{u}(t)\|_2\}\right]_k}}{\sqrt{N}},
  \label{eq:encoding}
\end{equation}
where $\mathcal{F}\{\cdot\}$ denotes the discrete Fourier transform over the spatial grid index, the overline denotes complex conjugation, and $|k\rangle$ is the qubit register state encoding Fourier mode index $k$. Because $\mathbf u(t)$ is a real-valued physical field, its Fourier-space coefficients satisfy the Hermitian-symmetry relation $\hat\psi_{N-k}=\overline{\hat\psi_k}$ for $k=1,\ldots,N/2-1$, while the zero-frequency and Nyquist coefficients are self-conjugate. Because $\hat\psi_k(t)$ is built directly from the \emph{normalized} field $\mathbf{u}(t)/\|\mathbf{u}(t)\|_2$, the register always stores a unit-norm Fourier-space state; no separate division by a norm is needed at the state-vector level. The physical norm $\|\mathbf{u}(t)\|_2$ is not itself represented on the register: it is stored and propagated entirely in classical memory alongside the quantum state and multiplied back after quantum readout. This separation is
essential because every quantum gate is unitary and
therefore norm-preserving, whereas the physical PDE
dynamics include dissipation and source injection that
alter the global amplitude.

The right-hand side of Eq.~\eqref{eq:semidiscrete} decomposes naturally into three physically distinct contributions. The diffusion operator is implemented in Fourier space via an ancilla-assisted controlled rotation (Section~\ref{sec:dad}). The advection operator is local in physical space and is realized by the Gray-code Trotter circuit (Section~\ref{sec:dad}). The forcing term $\mathbf{f}$ is a known, time-independent vector; because it requires no coherent superposition over unknown states, it is incorporated through a purely classical arithmetic update surrounding each quantum step (Section~\ref{sec:strang}). This division between quantum subroutines for homogeneous advection-diffusion dynamics and classical arithmetic for source injection is the backbone of the present method's hybrid architecture~\cite{gleinig2021efficient}. Within the homogeneous step, diffusion is applied as two symmetric half-steps flanking a single, unsplit advection step, a composition denoted diffusion-advection-diffusion (D--A--D) and detailed in Section~\ref{sec:dad}.

\subsection{Strang Splitting and Classical Source Integration}
\label{sec:strang}
The full one-step evolution is organized through symmetric Strang source splitting~\cite{strang1968construction}. Denoting the half-step source map as $\mathcal{S}(\Delta t/2)\colon \mathbf{u} \mapsto \mathbf{u} + (\Delta t/2)\,\mathbf{f}$ and the homogeneous advection-diffusion propagator over one step as $\mathcal{U}_{\mathrm{DAD}}(\Delta t)$, the update reads
\begin{equation}
  \mathbf{u}^{n+1}
  \;=\;
  \mathcal{S}\!\left(\tfrac{\Delta t}{2}\right)
  \circ\,
  \mathcal{U}_{\mathrm{DAD}}(\Delta t)
  \circ\,
  \mathcal{S}\!\left(\tfrac{\Delta t}{2}\right)
  \mathbf{u}^n,
  \label{eq:strang}
\end{equation}
Within the homogeneous propagator, diffusion is likewise split into two symmetric half-steps flanking a single, unsplit advection step,
$\mathcal{U}_{\mathrm{DAD}}(\Delta t) = \mathcal{D}(\Delta t/2)\circ\mathcal{A}(\Delta t)\circ\mathcal{D}(\Delta t/2)$
(Section~\ref{sec:dad}); source injection and diffusion are each applied as symmetric half-steps, while advection is applied once, unsplit, per iteration.
which is second-order accurate in time; the symmetric composition cancels the leading odd-order splitting error, giving a global error of $\mathcal{O}(\Delta t^2)$. The source half-steps are executed entirely in classical memory. At the beginning of each Strang step, the first classical source half-step gives
\begin{equation}
  \mathbf{v}_1 = \mathbf{u}^n + \tfrac{\Delta t}{2}\,\mathbf{f}.
  \label{eq:src_step1}
\end{equation}
The vector $\mathbf{v}_1$ is then normalized, transformed to the Fourier basis, and supplied to the quantum circuit  (Sections~\ref{sec:sparse_prep} and~\ref{sec:dad}), which produces the homogeneous solution $\mathbf{u}^{*}$. A second classical source half-step completes the time update.

\begin{equation}
  \mathbf{u}^{n+1} = \mathbf{u}^{*} + \tfrac{\Delta t}{2}\,\mathbf{f}.
  \label{eq:src_step2}
\end{equation}
Source injection is deterministic and requires neither ancilla qubits, post-selection, nor amplitude amplification; consequently
$P_{\mathrm{succ}}^{\mathrm{src}}=1$,
by construction. This value is not a probabilistic success rate but reflects the absence of any quantum post-selection step: source injection is a deterministic classical update, not a quantum circuit outcome.

After the first half-step source update, $\mathbf{v}_1$ is normalized and Fourier-transformed classically to yield the Fourier coefficient vector $\hat{\mathbf{S}} = \overline{\mathcal{F}\{\mathbf{v}_1/\|\mathbf{v}_1\|_2\}}/\sqrt{N}$. The norm $\|\mathbf{v}_1\|_2$ is stored classically and multiplied back after quantum readout to recover the physical field amplitude. In a prospective hardware or shot-based realization, where the post-selected outcome is itself a normalized state, this reconstruction additionally requires multiplying by $\sqrt{P_n}$, the classically tracked D-A-D success probability of Eq.~\eqref{eq:Pstep}; the statevector simulations reported here retain this factor automatically by working with the un-normalized post-selected branch. . The coefficient vector $\hat{\mathbf{S}}$ is then passed to the sparse Fourier state-preparation primitive (Section~\ref{sec:sparse_prep}), which encodes it directly into the quantum register in the Fourier basis, ready for the homogeneous quantum step.
\subsection{ Sparse Fourier State Preparation}
\label{sec:sparse_prep}
Before each application of the homogeneous quantum circuit, the data register is initialized with the normalized Fourier-space representation of the current field. Since the homogeneous evolution (Section~\ref{sec:dad}) begins with a diffusion operator that is diagonal in Fourier space, the state is prepared directly in the Fourier basis. This eliminates a redundant initial QFT and reduces the per-step gate cost by one $\mathcal{O}(n_q^2)$ operation.

After the first classical source half-step, $\mathbf{v}_1$ is normalized and Fourier-transformed classically to yield the target Fourier coefficient vector,
\begin{equation}
  \hat{S}_{k}
  = \frac{\overline{\left[\mathcal{F}\left\{\mathbf{v}_1/
    \|\mathbf{v}_1\|_2\right\}\right]_k}}{\sqrt{N}},
  \qquad k = 0,\ldots,N-1,
  \label{eq:Shat}
\end{equation}
where $\mathcal{F}\{\cdot\}$ denotes the discrete Fourier transform over the spatial grid index $m = 0,\ldots,N-1$, the overline denotes complex conjugation, and $k = 0,\ldots,N-1$ is the Fourier mode index. The complex conjugate and the $1/\sqrt{N}$ normalization ensure that the classically computed Fourier coefficients match the Fourier basis convention of the quantum circuit~\cite{qiskit2024}. The target quantum state is therefore
\begin{equation}
  |\hat{S}\rangle
  = \sum_{k=0}^{N-1} \hat{S}_{k}\,|k\rangle,
  \label{eq:Fourier_state}
\end{equation}
which resides directly in the Fourier-coefficient basis.

At step $n$, let $\mathcal I_n=\{k:|\hat S_k|>\delta\}$, $K_n=|\mathcal I_n|$, and $\eta_n=\|(\hat S_k)_{k\notin\mathcal I_n}\|_2$. The preparation uses the binary-tree state-preparation algorithm of Gleinig and Hoefler~\cite{gleinig2021efficient} (building on precursor amplitude-loading schemes for structured distributions~\cite{grover2002creating}), which loads the $K_n$ retained modes via a balanced binary cascade of controlled-$R_y$ rotations and phase gates.
Both the rotation angles and the phases of $\hat{S}_{k}$ are computed classically and encoded in the circuit. For the specific binary-tree construction considered here, the logical controlled-operation count scales as $\mathcal{O}(K_n n_q)$~\cite{mottonen2004}; the corresponding CX count depends on the hardware basis-gate decomposition under the fixed ancilla budget. The resulting advantage therefore applies when $K_n\ll N$, while the worst-case scaling is $\mathcal{O}(Nn_q)$.
Smoothness alone does not guarantee sparsity at a fixed absolute threshold. Indeed, the recorded $N=16$ viscosity runs have $K_n=N$ at the final step for $\delta=10^{-10}$. We therefore retain $K_n$ explicitly in every end-to-end resource estimate rather than assigning the two-mode support of the forcing to the source-corrected state.

The output state $|\hat{S}\rangle$ enters the homogeneous quantum step formulated directly in the Fourier basis; no further basis change is needed before the first diffusion substep (Fig.~\ref{fig:diffusion}(a)).

\subsection{ Homogeneous Quantum Step: Diffusion-Advection-Diffusion}\label{sec:dad}
The homogeneous quantum evolution applies the advection-diffusion operator $\mathcal{U}_{\mathrm{DAD}}(\Delta t)$ through a symmetric diffusion-advection-diffusion (D--A--D) splitting. The two operators require distinct quantum treatments: diffusion is non-unitary and is implemented using ancilla-assisted post-selection, whereas the advection operator is unitary and acts directly on the data register without ancillary qubits.

\subsubsection{Diffusion in Fourier Space}

The diffusion operator $\mathbf{D}_{\mathrm{diff}}$ is diagonal in Fourier space, so each Fourier mode $k$ evolves independently under the damping factor
\begin{equation}
  \lambda_k = \exp\!\left(-\nu \kappa_k^2 \frac{\Delta t}{2}\right),
  \label{eq:lambda}
\end{equation}
for a half-step of duration $\Delta t/2$.
Being a non-unitary operation, the diffusion half-step is applied within an ancilla-assisted uniformly controlled rotation (UCRY block)~\cite{harrow2009quantum,childs2012lcu,an2023linear}. With the data
register in the Fourier-basis superposition $\sum_k c_k |k\rangle$ and
an ancilla qubit initialized to $|0\rangle_{\mathrm{anc}}$, the joint
transformation is
\begin{equation}
  |k\rangle|0\rangle_{\mathrm{anc}}
  \;\longmapsto\;
  |k\rangle\!\left[
    \lambda_k\,|0\rangle_{\mathrm{anc}}
    + \sqrt{1-\lambda_k^2}\,|1\rangle_{\mathrm{anc}}
  \right],
  \label{eq:ucry}
\end{equation}
This operation is implemented by a mode-dependent $R_y(\theta_k)$ rotation with $\theta_k = 2\arccos(\lambda_k)$, conditioned on the $n_q$-qubit Fourier-mode index $|k\rangle$. The QFT and IQFT circuits are implemented using Qiskit's \texttt{QFTGate} primitive, whose canonical decomposition already includes the bit-reversal SWAP network required to produce natural-order output; no additional swap layer is appended in our implementation, so the native computational-basis index of the QFT/IQFT output coincides with the natural (non-bit-reversed) Fourier-mode order. The rotation angles $\theta_k$ are constructed directly from the natural-order signed wavenumbers of Eq.~\eqref{eq:signed_wavenumbers} and supplied to the UCRY block in the same natural index order $k=0,\ldots,N-1$, so no additional classical reordering of the angle table is required. Consistency between the QFT/IQFT output and the UCRY angle ordering therefore does not rely on any wire reordering internal to the UCRY block, but on pairing Qiskit's default swap-included QFT synthesis with a natural-order angle table. Post-selecting on $|0\rangle_{\mathrm{anc}}$ recovers the desired attenuated amplitude $\lambda_k c_k$ at mode $k$. By acting simultaneously on all populated modes in superposition, the UCRY block applies the full diagonal dissipation map within a single multiplexed controlled-rotation block. The rotation angles $\theta_k$ depend only on $\nu$, $k$, and $\Delta t$, all of which are fixed parameters, so the UCRY gate is pre-built once and reused at every time step (Fig.~\ref{fig:diffusion}(b)).

The post-selection probability for a normalized input state
$\sum_k c_k |k\rangle$ is
\begin{equation}
  P_{\mathrm{succ}}^{\mathrm{diff}}
  = \sum_k |c_k|^2\,\lambda_k^2.
  \label{eq:Psucc}
\end{equation}
For physically relevant parameters with small $\nu$ and $\Delta t$, we have $\lambda_k \approx 1$ for all modes, so $P_{\mathrm{succ}}^{\mathrm{diff}}$ remains close to unity. The quantitative values for the parameters used in the present study are reported in Section~\ref{sec:results}. In a prospective hardware implementation, post-selection failures can be handled via a per-step restart-until-success strategy, analogous to repeat-until-success (RUS) protocols~\cite{paetznick2013repeat}: upon an unsuccessful ancilla outcome (the UCRY measurement returning $|1\rangle_{\mathrm{anc}}$), the attempted step is discarded and the known input state is re-prepared from the classical checkpoint before the step is repeated. The abbreviation RUS is used subsequently to refer to this restart-based strategy. This avoids the multiplicative accumulation of post-selection probability over the full trajectory. For an individual diffusion half-step, the corresponding restart factor is $1/P_{\mathrm{succ}}^{\mathrm{diff}}$, whereas the success probability of a complete D-A-D step is the composite probability $P_n$ defined in Eq.~\eqref{eq:Pstep}.

\begin{figure}[t]
  \centering
  \includegraphics[width=\columnwidth]{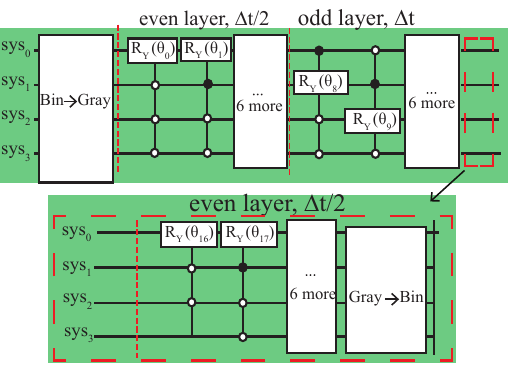}
  \caption{\leftskip0pt\rightskip0pt plus0pt\parfillskip0pt plus1fil\relax
    Quantum circuit for the advection substep
    $\mathcal{U}_{\mathrm{adv}}(\Delta t)$ via the
    Gray-code Trotter decomposition (Eq.~\ref{eq:adv_trotter}).
    The green background identifies this circuit as the
    detailed contents of the green $\mathrm{Adv}(\Delta t)$
    box in Fig.~\ref{fig:diffusion}. The circuit is shown in a
    two-row
    layout with vertical tick marks marking the row break.
    Read left to right, top row: \textbf{Bin$\to$Gray}, a
    classically precomputed permutation converting the
    register $sys_0$--$sys_3$ to Gray-code order so that
    adjacent grid indices $m$, $m+1$ differ in exactly one
    qubit; \textbf{even layer, $\Delta t/2$}, $(n_q-1)$-controlled
    $R_y(\theta)$ gates on the even edges $(m, m+1)$,
    $m = 0, 2, 4, \ldots$, each targeting the single qubit
    that differs between the two coupled Gray-code labels
    (filled/open control circles denote \texttt{ctrl}$=1$/$0$);
    representative gates $\theta_0$ (edge $(0,1)$, all
    anti-controls) and $\theta_1$ (edge $(2,3)$) are shown,
    with the dashed box standing for the remaining six;
    \textbf{odd layer, $\Delta t$}, the same construction on
    odd edges $m = 1, 3, 5, \ldots$, with representative
    gates $\theta_8$ (edge $(1,2)$) and $\theta_9$ (edge
    $(3,4)$).
    Bottom row: \textbf{even layer, $\Delta t/2$} repeats the
    even-edge layer at half step ($\theta_{16}$, $\theta_{17}$
    on the same edges as $\theta_0$, $\theta_1$), completing
    the symmetric splitting; \textbf{Gray$\to$Bin} restores
    binary ordering.
    The even-odd-even structure gives a local Trotter
    error of $\mathcal{O}(\Delta t^3)$, matching the Strang
    scheme's second-order accuracy. For $n_q = 4$ ($N = 16$),
    each layer has $N/2 = 8$ controlled $R_y$ gates --
    $3 \times 8 = 24$ total, each with $n_q - 1 = 3$ controls.
  }
  \label{fig:advection}
\end{figure}
\subsubsection{Advection via Gray-Code Trotter Decomposition}
 The advection operator $\mathbf{K}_{\mathrm{adv}}$ is local in physical space, coupling only nearest-neighbor grid points through the centered-difference stencil. Its skew-Hermitian structure ensures $\mathcal{U}_{\mathrm{adv}} = e^{\mathbf{K}_{\mathrm{adv}}\Delta t}$ is exactly unitary. Despite this structural simplicity, direct implementation in standard binary ordering is circuit-inefficient: adjacent grid points $m$ and $m+1$ can differ by multiple bits in their binary representations, requiring multi-qubit controlled gates for each coupling. For example, grid points $m=1$ (binary \texttt{01}) and $m=2$ (binary \texttt{10}) differ in two bits simultaneously, which would require a two-qubit controlled gate just for one nearest-neighbor coupling.

Gray-code ordering removes this inefficiency by relabeling the grid points according to the binary-reflected Gray code, $g(m)=m\oplus(m\gg1)$, for which adjacent labels differ by a single bit. For example, with $n_q=2$, the binary transition $01\rightarrow10$ flips two bits, whereas the Gray-code sequence $00\rightarrow01\rightarrow11\rightarrow10$ requires only one bit flip between successive states. The complete $n_q=2$ mapping is provided in Appendix~\ref{app:graycode}, Table~\ref{tab:gray_example}.

{Under this relabeling, each nearest-neighbor coupling becomes a transition on a single target qubit, with the remaining $n_q-1$ qubits defining the control pattern.} The logical transition therefore involves only one target qubit. Still, the basis-gate cost of the corresponding multicontrolled operation depends on the available work ancillas and hardware connectivity. The resulting circuit topology directly reflects the sparsity structure of the advection operator (Fig.~\ref{fig:advection}).

The nearest-neighbor edges of the advection operator are partitioned into two groups: even edges connecting pairs $(m, m+1)$ for
$m = 0, 2, 4, \ldots$ and odd edges connecting pairs $(m, m+1)$ for
$m = 1, 3, 5, \ldots$. Applying the symmetric second-order decomposition of Suzuki~\cite{suzuki1990fractal}, the full advection update over one time step is assembled as
\begin{equation}
  \mathcal{U}_{\mathrm{adv}}(\Delta t)
  \approx
  \mathcal{U}_{\mathrm{even}}\!\left(\tfrac{\Delta t}{2}\right)
  \circ\,
  \mathcal{U}_{\mathrm{odd}}(\Delta t)
  \circ\,
  \mathcal{U}_{\mathrm{even}}\!\left(\tfrac{\Delta t}{2}\right),
  \label{eq:adv_trotter}
\end{equation}
where $\mathcal{U}_{\mathrm{even}}$ and $\mathcal{U}_{\mathrm{odd}}$ are sequential applications of the controlled $R_y$ gates over the even and odd edge sets respectively. The rotation angle for edge $(m,m+1)$ is set as $\theta_m=-2\,(\mathbf K_{\rm adv})_{m,m+1}\,\tau$, where $(\mathbf K_{\rm adv})_{m,m+1}$ is the corresponding centered-difference off-diagonal coefficient and $\tau$ equals $\Delta t/2$ for the even layers and $\Delta t$ for the odd layer above. The symmetric composition yields a local splitting error of $\mathcal{O}(\Delta t^3)$, giving second-order accuracy in time. The combination of Gray-code ordering and symmetric Trotter decomposition is what makes the advection circuit both gate-efficient and second-order accurate: Gray-code ordering ensures each edge requires only one target qubit, while the symmetric Trotter sequence ensures the overall splitting error matches the second-order accuracy of the Strang scheme. The resulting circuit structure is shown in Fig.~\ref{fig:advection} for $n_q = 4$.

The complete homogeneous quantum update combines the preceding operators through a symmetric D--A--D composition. Let $F$ map position-space amplitudes to the Fourier convention used by the circuit. For an input state prepared in the Fourier basis, the successful branch is
\begin{align}
 |\psi_{\mathrm{out}}\rangle
 ={}&F^\dagger \mathcal D_{\frac12}
   F\mathcal A_2(\Delta t)
   F^\dagger \mathcal D_{\frac12}
   |\hat S\rangle , \notag\\
 \mathcal D_{\frac12}
 ={}&\operatorname{diag}
 \!\left(e^{-\nu\kappa_k^2\Delta t/2}\right),
 \label{eq:step_fourier}
\end{align}
where the operators act from right to left and $\mathcal A_2$ denotes the even-odd-even approximation of
Eq.~\eqref{eq:adv_trotter}. Since $|\hat S\rangle$ is prepared directly in the Fourier basis, no initial QFT is required.
The post-selection probability for one complete D--A--D step can be written as
\begin{equation}
  P_n=p_{1,n}\,p_{2,n|1},
  \label{eq:Pstep}
\end{equation}
where $p_{1,n}$ is the success probability of the first diffusion half-step and $p_{2,n|1}$ is that of the second, conditioned on successful completion of the first. The relation $P_n=p_{1,n}^2$ holds only when the spectral distributions entering the two diffusion steps are identical.

\subsection{Accuracy decomposition and commutator-based error estimate}
\label{sec:accuracy_theory}
The comparisons below use the adaptive RK45 scheme and the explicit eighth-order DOP853 method, both applied to the semi-discrete system in Eq.~\eqref{eq:semidiscrete}. Because these classical solutions retain the same spatial discretization, they serve as temporal-integration references rather than continuum benchmarks. A consistent error assessment therefore requires separating spatial discretization and temporal integration errors. For smooth periodic fields $u$ and $U$, a Taylor expansion of Eq.~\eqref{eq:Kadv} gives
\begin{equation}
 \mathbf K_{\mathrm{adv}}\mathbf P_hu
 =\mathbf P_h\!\left[-\frac12
 (Uu_x+(Uu)_x)\right]+\mathcal O(h^2),
 \label{eq:spatial_consistency}
\end{equation}
where $\mathbf P_h$ samples the continuum field. The
Fourier Laplacian is spectrally accurate for smooth
periodic fields, so the centered advection stencil sets
the algebraic spatial order.

To characterize the temporal error, we augment the state as
$\widetilde{\mathbf u}=(\mathbf u,1)^T$ and define the corresponding generators
\begin{align}
 \widetilde S&=\begin{pmatrix}0&\mathbf f\\0&0\end{pmatrix},
 &\widetilde D&=\begin{pmatrix}\mathbf D_{\rm diff}&0\\0&0\end{pmatrix},
 \notag\\
 \widetilde K_e&=\begin{pmatrix}\mathbf K_e&0\\0&0\end{pmatrix},
 &\widetilde K_o&=\begin{pmatrix}\mathbf K_o&0\\0&0\end{pmatrix},
 \label{eq:augmented_generators}
\end{align}
where $\mathbf K_e+\mathbf K_o=\mathbf K_{\rm adv}$ is
the even--odd edge partition. For a symmetric pair
$e^{\tau A/2}e^{\tau B}e^{\tau A/2}$, we define
\begin{equation}
 c(A,B)=\frac{\|[A,[A,B]]\|_2}{24}
       +\frac{\|[B,[B,A]]\|_2}{12}.
 \label{eq:commutator_proxy}
\end{equation}
Denote by $\Phi_{\Delta t}$ the one-step Trotterized propagator
composed of the three symmetric substeps above, and define
$\widetilde L_h=\widetilde S+\widetilde D+\widetilde K$, with
$\widetilde K=\widetilde K_e+\widetilde K_o$. The augmented semidiscrete
system then satisfies $\partial_t\widetilde{\mathbf u}=\widetilde
L_h\widetilde{\mathbf u}$, so $e^{T\widetilde L_h}$ is the exact augmented
semidiscrete flow over the interval $T$. Applying the Baker--Campbell--Hausdorff expansion~\cite{childs2021trotter} to the
three nested symmetric compositions, we obtain the global
semidiscrete estimate of
\begin{align}
 \|\Phi_{\Delta t}^{M}-e^{T\widetilde L_h}\|_2
 \leq{}&T\Delta t^2 e^{T\Lambda_h}
 \big[c(\widetilde S,\widetilde D+\widetilde K)
 +c(\widetilde D,\widetilde K) \notag\\
 &+c(\widetilde K_e,\widetilde K_o)\big]
 +\mathcal O(T\Delta t^4),
 \label{eq:global_error_bound}
\end{align}
where $M=T/\Delta t$, and
$\Lambda_h=
\|\widetilde S\|_2
+\|\widetilde D\|_2
+\|\widetilde K_e\|_2
+\|\widetilde K_o\|_2$ is the sum of generator  norms. The deliberately conservative leading-order commutator estimate makes the dependence of the error constant on velocity, viscosity, grid spacing, and forcing explicit.

If Fourier truncation is used in state preparation, its
normalized one-step error is at most
$2\eta_n/\sqrt{1-\eta_n^2}$. Stability then adds the
term
\begin{equation}
 e^{T\Lambda_h}\sum_{n=0}^{M-1}
 \frac{2\eta_n}{\sqrt{1-\eta_n^2}}
 \label{eq:prep_error_bound}
\end{equation}
to Eq.~\eqref{eq:global_error_bound}. Conditional on
successful ancilla outcomes, the ideal UCRY diffusion
map itself is exact; post-selection changes cost, not
algorithmic bias. The accompanying analysis script
evaluates Eqs.~\eqref{eq:spatial_consistency}--\eqref{eq:global_error_bound}
and the observed local $\mathcal O(\Delta t^3)$ defect. Because
$\|\mathbf D_{\rm diff}\|_2$ grows as $N^2$, the asymptotic temporal audit uses
$\Delta tN^2=\mathrm{constant}$ under grid refinement; a fixed $\Delta t$ can
otherwise leave the small-step regime on the finer grids.

\subsection{Qubit Budget and Circuit Cost}
\label{sec:cost}

The homogeneous quantum kernel requires $n_q$ data
qubits plus one diffusion ancilla. The two UCRY
applications share that ancilla through mid-circuit
measurement and reset, so the peak logical count is
\begin{equation}
  n_{\mathrm{peak}} = n_q + 1.
  \label{eq:nqubits}
\end{equation}
No additional logical qubits are used by the classical source update or the ancilla-free binary-tree preparation. This peak count does not, however, determine the gate or sampling complexity.

For simulation convenience, the statevector validation reported in
Section~\ref{sec:results} represents the two diffusion half-steps using
separate ancilla registers rather than performing an explicit mid-circuit
measurement and reset.  In the intended dynamic-circuit implementation, a single diffusion ancilla is measured, reset, and reused between the two half-steps, yielding the peak logical-qubit requirement
$n_q+1$ given in Eq.~\eqref{eq:nqubits}.

One successful kernel call contains $3N/2$
$(n_q-1)$-controlled $R_y$ rotations for advection,
two $N$-mode UCRY blocks, and three Fourier transforms~\cite{coppersmith1994approximate}.
The corresponding logical block counts therefore scale as
\begin{equation}
 G_{\rm kernel}
 =\frac{3N}{2}G_{\rm mc}(n_q-1)
  +\mathcal O(N)+\mathcal O(n_q^2),
 \label{eq:kernel_cost}
\end{equation}
where $G_{\rm mc}$ is the basis-gate cost of one multicontrolled rotation. It depends on the synthesis model: connectivity, clean work ancillas, and relative-phase constructions must be specified. Consequently, $3N/2$ is a hardware-independent logical rotation count, whereas an unconditional $\mathcal O(Nn_q)$ CX claim is not justified under the fixed $n_q+1$-qubit budget.

The end-to-end hybrid step additionally requires a classical FFT, preparation of the actual source-corrected state at $\mathcal O(K_n n_q)$ gates, an expected $1/P_n$ kernel attempts, and phase-sensitive reconstruction of the output amplitudes. Computational-basis sampling alone returns $|u_m|^2$ and cannot supply the signs or phases needed for the next classical source addition. Even a probability histogram has
\begin{equation}
 \mathbb E[\mathrm{TVD}]
 \leq \frac12\sqrt{\frac{N-1}{N_s}},
 \label{eq:shot_bound}
\end{equation}
while generic amplitude reconstruction requires at
least order $N/\epsilon_{\rm meas}^2$ state copies. For
$M$ steps, a transparent cost model is therefore
\begin{align}
  C_{\rm end}
 =\sum_{n=0}^{M-1}\big[
  &(G_{\rm prep}(K_n)+G_{\rm kernel})/P_n \notag\\
  &+C_{\rm tomo}(N,\epsilon_{\rm meas})
  +C_{\rm FFT}(N)\big].
 \label{eq:end_to_end_cost}
\end{align}

We report the transpiled homogeneous kernel separately
from these end-to-end terms in Sec.~\ref{sec:res_cost}.

\raggedbottom

\begin{table}[H]
\centering
\caption{\leftskip0pt\rightskip0pt plus0pt\parfillskip0pt plus1fil\relax Validation at $\Delta t = 10^{-3}$, $T = 1.0$. Relative $L_2$ errors
of the quantum solution against the matched split-step reference (isolating the Trotter error of the advection circuit) and against DOP853 (total temporal
error), together with the measured post-selection statistics of the diffusion ancilla. The source steps succeed with probability exactly one in
all $2000$ applications.}
\label{tab:validation}
\begin{tabular}{cccccc}
\hline
$n_q$ & $N$ & $L_2$ vs.\ matched & $L_2$ vs.\ DOP853 &
$\bar{P}_{\rm succ}$ & $\min P_{\rm succ}$ \\
\hline
4 & 16 & $2.184\times10^{-5}$ &
$2.232\times10^{-5}$ & $0.999349$ &
$0.998896$ \\
5 & 32 & $1.012\times10^{-4}$ &
$1.015\times10^{-4}$ & $0.999369$ &
$0.998892$ \\
\hline
\end{tabular}
\end{table}

\section{Results}
\label{sec:results}

Unless stated otherwise, all results use the configuration defined in Section~\ref{sec:governing}: $U(x) = 1.0 + 0.3\sin(2\pi x/L)$, $\nu = 0.008$,
$f(x) = 0.5\sin(2\pi x/L)$, a normalized Gaussian initial condition,
$\Delta t = 10^{-3}$, and $T = 1.0$, with
$N=16$ ($n_q=4$) and $N=32$ ($n_q=5$).
All quantum calculations employ the Gray-code Trotterized advection circuit and the classical Strang source update.
The circuits are simulated at the statevector level in Qiskit~\cite{qiskit2024} to
isolate algorithmic error, while finite-shot sampling effects and a
depolarizing-plus-readout noise model are considered in
Section~\ref{sec:res_shots} and Appendix~\ref{app:noise}.

\subsection{Validation against classical references}
\label{sec:res_validation}

Three classical references with distinct roles are used. The matched split-step solver applies the identical Strang composition and identical discrete
operators in floating-point arithmetic, with the single exception that the advection exponential $e^{\mathbf{K}_{\rm adv}\Delta t}$ is evaluated exactly.
Its difference from the quantum solution therefore isolates the Gray-code Trotter error, up to the independently controlled state-preparation truncation error; conditioned on successful post-selection, the UCRY diffusion and QFT operations are algebraically exact.

\begin{figure}[t]
\centering
\includegraphics[width=\linewidth]{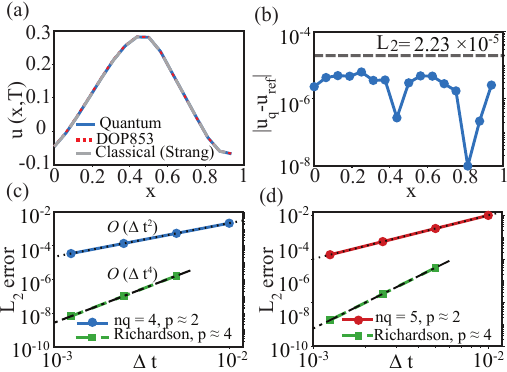}
\captionsetup{
    width=\linewidth,
    singlelinecheck=false
}
\caption{\protect \justifying Validation and temporal convergence. (a)~Quantum field at
$T = 1.0$, $\Delta t = 10^{-3}$, $n_q = 4$, against the DOP853 and matched
split-step references. (b)~Pointwise error against the DOP853 reference on a logarithmic scale; the matched split-step error curve is indistinguishable at this scale. (c),(d)~Temporal convergence against DOP853 for
$n_q = 4$ and $n_q = 5$: the base scheme follows the second-order guide line over the full $\Delta t$ range, while the Richardson-extrapolated solutions
lie two to four orders of magnitude below it and converge at fourth order.
The corresponding $n_q = 5$ validation panels appear in
Appendix~\ref{app:robustness}.}
\label{fig:validation}
\end{figure}

\begin{table*}[t!]
\centering
\caption{\leftskip0pt\rightskip0pt plus0pt\parfillskip0pt plus1fil\relax
Temporal convergence against DOP853 and Richardson extrapolation
over the pair $(\Delta t, 2\Delta t)$. The observed order $p = \log_2[e(2\Delta t)/e(\Delta t)]$ confirms second-order accuracy;
successive Richardson errors decrease by factors of $16.0$, identifying
fourth-order behavior of the extrapolated solution.}
\label{tab:convergence}
\begin{tabular}{ccccccc}
\hline
& \multicolumn{3}{c}{$n_q = 4$} & \multicolumn{3}{c}{$n_q = 5$} \\
$\Delta t$ & $L_2$ error & $p$ & Richardson & $L_2$ error & $p$ & Richardson \\
\hline
$0.01$    & $2.23\times10^{-3}$ & --      & --                  & $1.01\times10^{-2}$ & --      & -- \\
$0.005$   & $5.58\times10^{-4}$ & $2.000$ & $1.65\times10^{-6}$ & $2.54\times10^{-3}$ & $1.998$ & $4.04\times10^{-5}$ \\
$0.0025$  & $1.39\times10^{-4}$ & $2.000$ & $1.03\times10^{-7}$ & $6.34\times10^{-4}$ & $2.000$ & $2.52\times10^{-6}$ \\
$0.00125$ & $3.50\times10^{-5}$ & $2.000$ & $6.46\times10^{-9}$ & $1.59\times10^{-4}$ & $2.000$ & $1.57\times10^{-7}$ \\
\hline
\end{tabular}
\end{table*}

Two further numerical checks accompany this comparison. The first
audits the encoded real-valued initial condition against the
Hermitian-symmetry relation of Sec.~\ref{sec:governing}: the maximum
symmetry residual, $\max_k|\hat\psi_{N-k}-\overline{\hat\psi_k}|$ over
$k=1,\ldots,N/2-1$, was $4.1\times10^{-16}$ for $n_q=4$ and
$6.4\times10^{-16}$ for $n_q=5$, with the imaginary parts of the
zero-frequency and Nyquist coefficients below $10^{-17}$. This
supports, but does not by itself demonstrate, Hermitian symmetry
at later evolution times. The second compares the complete QFT-UCRY-IQFT diffusion
block with direct classical spectral damping $\hat
u_k\mapsto\lambda_k\hat u_k$ at the canonical parameters $n_q=4$,
$\nu=0.008$, and $\Delta t=10^{-3}$, where the maximum pointwise amplitude
residual between the normalized post-selected quantum output and the
normalized classically damped state was found to be $8.6\times10^{-16}$, confirming
consistency of the QFT convention, mode ordering, and UCRY angle
assignment to machine precision.

Two independent adaptive integrations, RK45 and DOP853, use a
relative tolerance of $10^{-11}$ to integrate the same semi-discrete system,
Eq.~\eqref{eq:semidiscrete}, in continuous time, so the deviation from them
measures the total temporal discretization error. Their relative $L_2$
distance is $5.97\times10^{-12}$ for $N=16$ and
$1.02\times10^{-11}$ for $N=32$, which certifies that the semi-discrete
reference introduces no ambiguity at the accuracy levels reported below.

Table~\ref{tab:validation} and Figs.~\ref{fig:validation}(a,b) summarize the
outcome. The errors against the matched split-step and DOP853
references are close in magnitude, indicating that Gray-code Trotter
splitting provides the dominant contribution to the measured temporal error
at this time step, while the Strang source splitting contributes negligibly. The factor $\simeq 4.6$
increase from $n_q = 4$ to $n_q = 5$ at fixed $\Delta t$  is consistent with growth of the even--odd commutators under grid refinement; the corresponding commutator-scaling estimate
is given by Eq.~\eqref{eq:global_error_bound}. This is purely a temporal-convergence comparison and the spatial-convergence check will be discussed next. The complete validation is reproduced with a
four-harmonic source $f(x)$ in Appendix~\ref{app:robustness}, showing the same convergence behavior with quantitatively similar error and success-probability levels.

\FloatBarrier
\subsection{Convergence: temporal and spatial}
\label{sec:res_convergence}

Table~\ref{tab:convergence} and Figs.~\ref{fig:validation}(c,d) establish two
results. First, the observed convergence order is approximately $2.000$ at both resolutions, confirming that the composition of Strang source
splitting, the D-A-D operator factorization, and the even-odd Trotterized advection retains clean second-order accuracy through the complete quantum
pipeline. Second, and less trivially, the Richardson-extrapolated solution
$(4u_{\Delta t} - u_{2\Delta t})/3$ does not merely gain one order: successive
extrapolated errors fall by factors of $16.0$ at both resolutions, i.e., the extrapolation converges at fourth order. This is a direct consequence of the
time-symmetric construction of every splitting layer~\cite{strang1968construction,suzuki1990fractal},
which restricts the error expansion to even powers of $\Delta t$, so that
eliminating the $\Delta t^2$ term exposes $\Delta t^4$. This reduction in error also lowers the number of ideal kernel evaluations required for a given accuracy. At $n_q = 4$,
the Richardson pair $(\Delta t, 2\Delta t) = (0.005, 0.01)$ requires $300$
ideal kernel calls and reaches an error of $1.65\times10^{-6}$, whereas the
unextrapolated scheme requires $800$ kernel calls at $\Delta t = 0.00125$
to reach only $3.5\times10^{-5}$. A fitted accuracy-to-kernel-cost
projection is reported in Appendix~\ref{app:accuracy_cost},
Table~\ref{tab:accuracy_cost_projection}. The projection indicates an
ideal-kernel-count reduction of approximately a factor of four at a target
error of $10^{-4}$ and approximately thirteen at $10^{-6}$. These savings
do not include state preparation, post-selection retries, or
reconstruction and therefore are not interpreted as an end-to-end quantum
speedup.

The convergence analysis above isolates the temporal error relative to the semi-discrete reference. We next examine spatial convergence and how the semi-discrete operator approaches the continuum PDE under grid refinement.

Table~\ref{tab:operator_audit} provides the continuum consistency check. The spatial refinement order approaches two
monotonically, while the local operator defect approaches
$\mathcal O(\Delta t^3)$ at every resolution once the diffusive scale is
resolved. Together with the global temporal slopes in
Table~\ref{tab:convergence}, these results distinguish the
$\mathcal O(h^2)$ spatial error from the $\mathcal O(\Delta t^2)$ accumulated
splitting error. The manufactured-solution columns confirm
this independently for the complete forced partial PDE rather than only the
semi-discrete generator. In particular, their refinement order approaches two, so both
audits agree that the centered advection stencil sets the spatial error
while the Fourier diffusion discretization remains more accurate for these
smooth periodic data.

\begin{table*}[t!]
\small
\centering
\caption{\leftskip0pt\rightskip0pt plus0pt\parfillskip0pt plus1fil\relax
Independent operator audit on grids from $N=8$ to $256$.
$E_h$ is the relative consistency error in
Eq.~\eqref{eq:spatial_consistency}, $p_h$ is its grid-refinement order, and
$p_{\rm loc}$ is the observed order of the one-step operator defect between
the split and exact augmented maps. The reported $p_{\rm loc}$ compares the
two finest steps with $\Delta tN^2=0.16$ and $0.08$. The
last two columns add the full-PDE manufactured-solution check at $T=0.1$
(exact solution $u=e^{-t}[\sin(2\alpha x)+0.2\cos(3\alpha x)]$, forcing
evaluated analytically for $U=1+0.3\sin(\alpha x)$, DOP853 tolerances
$10^{-11}$ and $10^{-13}$).}
\label{tab:operator_audit}
\begin{tabular}{ccccccc}
\hline
& & \multicolumn{3}{c}{semidiscrete operator audit} &
\multicolumn{2}{c}{full-PDE manufactured solution} \\
$n_q$ & $N$ & $E_h$ & $p_h$ & $p_{\rm loc}$ &
$L_2$ error & $p_h$ \\
\hline
3 & 8   & $3.7281\times10^{-1}$ & --       & $2.9971$ & -- & -- \\
4 & 16  & $1.0296\times10^{-1}$ & $1.8564$ & $2.9971$ & $1.14889\times10^{-1}$ & -- \\
5 & 32  & $2.6451\times10^{-2}$ & $1.9607$ & $2.9971$ & $2.91268\times10^{-2}$ & $1.97982$ \\
6 & 64  & $6.6584\times10^{-3}$ & $1.9901$ & $2.9971$ & $7.30389\times10^{-3}$ & $1.99561$ \\
7 & 128 & $1.6675\times10^{-3}$ & $1.9975$ & $2.9970$ & $1.82731\times10^{-3}$ & $1.99894$ \\
8 & 256 & $4.1704\times10^{-4}$ & $1.9994$ & $2.9969$ & $4.56911\times10^{-4}$ & $1.99974$ \\
\hline
\end{tabular}
\end{table*}
\subsection{State preparation accuracy}
\label{sec:res_prep}

Temporal and spatial discretization are not the only pipeline error sources: sparse Fourier state preparation introduces
an additional error through threshold-based Fourier-mode truncation. The
retained support $K_n$ depends on both the truncation threshold and the
evolving source-corrected state and therefore cannot be inferred from the
sparsity of the forcing alone. The detailed
minimum, median, and maximum $K_n$, discarded norms $\eta_n$, and
corresponding one-step truncation bounds are reported in
Appendix~\ref{app:prep_audit}, Table~\ref{tab:prep_audit}. The result shows
that tightening the threshold substantially increases the retained
support and that, at stringent tolerances, $K_n$ can approach $N$; at the
$10^{-10}$ threshold, the recorded $N=64$ trajectory reaches $K_n=N$.
Robustness was examined over twelve amplitude-wavenumber
parameter combinations spanning amplitude $\{0, 0.3, 0.6, 0.9\}$ crossed with
wavenumber $\{1, 2, 4\}$ in $U(x) = 1 + a\sin(2\pi\,k\, x/L)$; because
$a=0$ gives the same flat profile $U=1$ regardless of $k$, these
combinations span ten distinct velocity fields. The per-step diffusion
success probability generally decreases with velocity contrast
$v_{\max}-v_{\min}$, from $0.99904$ at zero contrast to a minimum of
$0.99596$ at the largest contrast tested ($U \in [0.1, 1.9]$, $k=4$). The
relative $L_2$ error follows a different pattern: at the fundamental
wavenumber ($k=1$) it grows mildly, whereas at the higher
wavenumbers $k=2$ and $k=4$ it instead falls, to as low as
$3.6\times10^{-5}$. The worst-case error is therefore specific to the
fundamental-mode, high-contrast corner of the sweep rather than typical of
the full grid. The skew-Hermitian defect of the audited advection matrices was
zero to machine precision, and the largest diffusion Hermiticity defect was
$1.65\times10^{-12}$.

\subsection{Energy budget and the ancilla--dissipation identity}
\label{sec:res_energy}

\begin{figure}[H]
\centering
\includegraphics[width=\columnwidth]{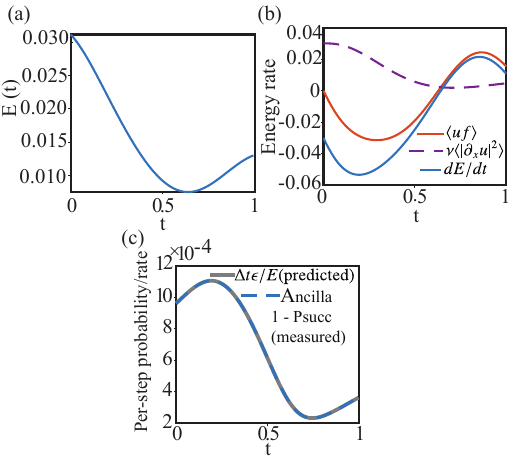}
\caption{Energy budget and the ancilla--dissipation identity at $n_q = 4$.
(a)~Total energy $E(t)$. (b)~Injection $\langle u f\rangle$, dissipation
$\nu\langle|\partial_x u|^2\rangle$, and $dE/dt$; the balance residual is
bounded by $1.2\times10^{-4}$, consistent with the exact discrete
energy conservation implied by the skew-Hermitian advection operator. (c)~Measured per-step ancilla loss
$1 - P_{\rm succ}$ against the step-centered numerical evaluation of the
leading-order relation in
Eq.~\eqref{eq:ancilla_dissipation}; the curves are indistinguishable
(correlation within $4\times10^{-9}$ of unity).}
\label{fig:energy}
\end{figure}

The physical content of the post-selection statistics
admits a sharp characterization. Expanding the UCRY
success probability of Eq.~\eqref{eq:Psucc} for small
$\nu\kappa_{\max}^2\Delta t$ and combining the two
diffusion half-steps yields
\begin{align}
  1-P_n&=\Delta t\,\frac{\varepsilon(t_n)}{E(t_n)}
  +\mathcal O\!\left[(\nu\kappa_{\max}^2\Delta t)^2+\Delta t^2+\eta_n\right],\notag\\
  E&=\tfrac{1}{2}\!\int u^2\,\dd x,\qquad
  \varepsilon=\nu\!\int |\partial_xu|^2\,\dd x,
  \label{eq:ancilla_dissipation}
\end{align}
Equation~\eqref{eq:ancilla_dissipation} is the analytical
leading-order relation. The numerical comparisons in Fig.~\ref{fig:energy}(c)
and Table~\ref{tab:postselection_remainder} evaluate it using a
step-centered estimate: writing $E_n=E(t_n)$ and $\varepsilon_n=\varepsilon(t_n)$,
with $\bar E_n=\tfrac12(E_n+E_{n+1})$ and
$\bar\varepsilon_n=\tfrac12(\varepsilon_n+\varepsilon_{n+1})$, the evaluated
prediction is $\Delta t\,\bar\varepsilon_n/\bar E_n$.
This identifies the per-step ancilla failure
probability with the instantaneous relative viscous
dissipation rate. For either half-step the scalar inequality
$0\leq x-(1-e^{-x})\leq x^2/2$ supplies a direct
remainder bound; the second $\mathcal O(\Delta t^2)$ term
accounts for the changed spectrum between the two
halves. Figure~\ref{fig:energy}
verifies this identity directly: the measured $1 - P_{\rm succ}$ and the
step-centered evaluation $\Delta t\,\bar\varepsilon_n/\bar E_n$ of the right-hand side of
Eq.~\eqref{eq:ancilla_dissipation}, evaluated from the
solution fields, have a correlation coefficient
within $4\times10^{-9}$ of unity across all $1000$ steps, reproducing
the full non-monotonic time history of the
dissipation. The correlation confirms the predicted
time dependence, while the remainder above is required
to assess magnitude agreement. The ancilla measurement
record can therefore act as a calibrated dissipation
diagnostic in the small-step regime.

The central panel of Fig.~\ref{fig:energy} additionally verifies the discrete energy budget $dE/dt = \langle u f\rangle - \varepsilon$: the residual is
bounded by $1.2\times10^{-4}$ against budget terms of order $3\times10^{-2}$, and is attributable to the $\mathcal{O}(\Delta t^2)$ splitting error and the
finite-difference evaluation of $dE/dt$. Advection contributes identically zero to the budget, which is the discrete statement that the exact unitarity
of the Gray-code circuit and the energy conservation of transport are the same property.

The magnitude, rather than only the correlation, of
Eq.~\eqref{eq:ancilla_dissipation} is tested separately by refining
$\Delta t$ and tracking the maximum ancilla--dissipation remainder; to keep
the main text focused on the primary results, that refinement study is
reported in Table~\ref{tab:postselection_remainder} in
Appendix~\ref{app:noise}.

\subsection{Post-selection statistics across viscosity}
\label{sec:res_psucc}

\begin{figure}[t]
\centering
\includegraphics[width=\linewidth]{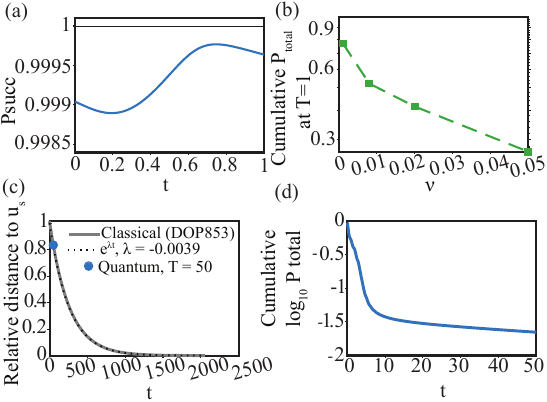}
\captionsetup{
    width=\linewidth,
    singlelinecheck=false
}
\caption{\protect \justifying Post-selection statistics across viscosity and horizon length.
(a)~Per-step success probability at the canonical viscosity $\nu = 0.008$,
recovering toward unity as dissipation empties the high-wavenumber modes.
(b)~ Cumulative success probability across the viscosity sweep of
Table~\ref{tab:nusweep}. (c)~Long-horizon integration to $T = 50$: relative
distance to the exact steady state $u_s = -\mathbf{A}^{-1}\mathbf{f}$,
where $\mathbf{A}$ is the discrete steady-state advection-diffusion
operator and $\mathbf{f}$ the discrete forcing vector, along
the classical trajectory, decaying on the slow-mode timescale $\tau = 257$;
the quantum solution at $T = 50$ (symbol) lies on the classical curve.
(d)~Cumulative $\log_{10}P$ of the post-selection record over the full
$5\times10^{4}$ steps, flattening as the spectrum concentrates in weakly
dissipated modes.}
\label{fig:psucc}
\end{figure}

Table~\ref{tab:nusweep} and Figs.~\ref{fig:psucc}(a,b) quantify the single failure channel of the algorithm across a fifty-fold range of viscosity, from $\nu = 0.001$ to $0.05$. Over this range, the cumulative success probability at $T = 1$ decreases from $0.773$ to
$0.268$, a factor of approximately $2.9$. The mechanism follows Eq.~\eqref{eq:ancilla_dissipation}: the post-selection loss is concentrated in the high-wavenumber modes, and viscosity removes precisely those modes, so the failure channel depletes its
own source. The per-step traces show the corresponding signature, with $P_{\rm succ}$ recovering toward unity as the spectrum empties. Under the Repeat-Until-Success protocol~\cite{paetznick2013repeat}, in which the ancilla is measured after each UCRY block and the step is re-prepared from the classically stored field upon failure, using $1/\bar P_{\rm succ}$ as a reciprocal-mean proxy, the retry overhead remains below $0.14\%$ per step throughout the sweep. Here, because
each retry restarts from the same classically stored input, it introduces no additional algorithmic approximation in the  model. Accuracy simultaneously improves with $\nu$,
as dissipation suppresses the high-$k$ content in which the Trotter commutator error resides. Thus, no adverse accuracy-success-probability tradeoff is observed over the tested parameter range.

\begin{table}[t!]
\centering
\caption{\leftskip0pt\rightskip0pt plus0pt\parfillskip0pt plus1fil\relax Viscosity sweep at $n_q = 4$, $\Delta t = 10^{-3}$, $T = 1.0$.
A fifty-fold increase in $\nu$ reduces the cumulative success probability by
only a factor of $2.9$, and the reciprocal-mean Repeat-Until-Success (RUS) calls/step proxy stays
below $1.0014$ (an overhead below $0.14\%$) per step throughout.}
\label{tab:nusweep}
\resizebox{\columnwidth}{!}{%
\begin{tabular}{cccccc}
\hline
$\nu$ & $L_2$ vs.\ DOP853 & $\bar{P}_{\rm succ}$ & $\min P_{\rm succ}$ &
$P_{\rm total}(T{=}1)$ & RUS calls/step proxy \\
\hline
$0.001$ & $2.9\times10^{-5}$ & $0.99974$ & $0.99964$ & $0.773$ & $1.0003$ \\
$0.008$ & $2.2\times10^{-5}$ & $0.99935$ & $0.99890$ & $0.522$ & $1.0007$ \\
$0.02$  & $1.6\times10^{-5}$ & $0.99913$ & $0.99760$ & $0.417$ & $1.0009$ \\
$0.05$  & $1.2\times10^{-5}$ & $0.99868$ & $0.99404$ & $0.268$ & $1.0013$ \\
\hline
\end{tabular}%
}
\end{table}

\subsection{Measurement statistics and source sparsity}
\label{sec:res_shots}

Sampling the final state with $N_s$ shots reproduces the expected statistical  behavior, with the total variation distance to the exact distribution falling
from $4.3\times10^{-2}$ at $10^3$ shots to $7.0\times10^{-3}$ at
$5\times10^{4}$ shots, consistent with the $N_s^{-1/2}$ scaling (Fig.~\ref{fig:shots}(a), Appendix~\ref{app:noise}). Repeating the complete simulation and
measurement with the four-harmonic source yields total variation distances that are equal to or smaller than the two-mode case at every shot count.

Forcing complexity does not introduce a separate probabilistic source step as shown in (Fig.~\ref{fig:shots}(b)).
No additional sampling penalty is observed for this particular two-mode versus four-harmonic comparison, though this conclusion is not established for arbitrary forcing bandwidth. It can instead impact the step-dependent preparation cost indirectly through the spectral support $K_n$ of the resulting source-corrected state, for which the stated binary-tree construction gives a logical preparation count scaling as $\mathcal{O}(K_n n_q)$. It can also affect
the diffusion post-selection probability indirectly by changing the evolved Fourier spectrum. In the rerun, the four-harmonic source contains eight active
conjugate Fourier modes, and its relative errors against DOP853 are
$2.57\times10^{-5}$ for $n_q=4$ and $1.29\times10^{-4}$ for $n_q=5$, with
mean per-step success probabilities $0.999247$ and $0.999234$, respectively.
Two further finite-shot results are deferred to
Appendix~\ref{app:noise},  where under the depolarizing-plus-readout noise model, the sampling-circuit distributional error varies only weakly with gate-error strength at fixed readout error, suggesting readout error contributes an important error floor. Also, we show that applying the Richardson extrapolation application to
measured probability distributions is beneficial only above a shot threshold at which sampling noise falls below the splitting error. Below this threshold, the extrapolation amplifies shot noise by the factor $\sqrt{17}/3 \approx 1.37$, in agreement with the measured degradation.

\subsection{Circuit resources}
\label{sec:res_cost}

\begin{figure}[H]
\centering
\includegraphics[width=\linewidth]{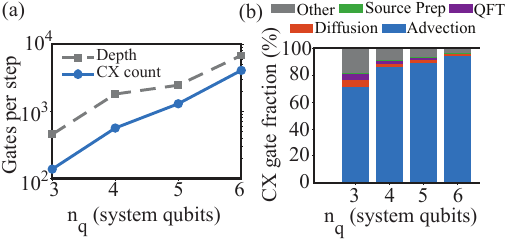}
\caption{Transpiled diagnostic of the homogeneous D-A-D quantum kernel for $n_q = 3$--$6$. (a)~Depth
and CX count per kernel. (b)~Stacked CX contribution by circuit block; the
Gray-code advection circuit, which encodes the $N$ values of $U(x)$,
dominates at all sizes.}
\label{fig:cost}
\end{figure}

Figure~\ref{fig:cost} summarizes the transpilation diagnostic.
Within this homogeneous circuit, the advection block contributes
$71\%$--$95\%$ of the CX count across the tested register sizes. The
underlying depth and gate counts are reported in
Appendix~\ref{app:resource_audit}, Table~\ref{tab:transpilation_audit}.
These counts do not establish an end-to-end Strang-step cost because state
preparation, source arithmetic, phase-sensitive reconstruction,
mid-circuit measurement/reset, and device routing are excluded.

Accordingly, the transpilation results indicate kernel-level scaling rather than end-to-end hardware cost or feasibility. A device-specific assessment would require transpilation of the complete Fourier-input dynamic circuit, including step-dependent state preparation, reuse and reset of the diffusion ancilla, and the target-device connectivity. Hardware-independent kernel and end-to-end resource estimates are given in Eqs.~\eqref{eq:kernel_cost} and~\eqref{eq:end_to_end_cost}, respectively.

Table~\ref{tab:resource_projection} makes the hybrid boundary quantitative.
The compact register grows only from five to thirteen logical qubits over the
displayed range, but the loaded velocity data, controlled rotations, and
measurement copies grow with $N$. At $n_q=12$, for example, a single
homogeneous kernel contains $6144$ logical advection rotations and $8192$
logical diffusion-mode rotations, while generic phase-sensitive
reconstruction at one-percent accuracy has a $4.096\times10^7$-copy proxy.
These figures support the few-step kernel target but not an end-to-end
exponential speedup for the present readout-and-repreparation workflow.

\begin{table*}[t]
\centering
\caption{\leftskip0pt\rightskip0pt plus0pt\parfillskip0pt plus1fil\relax
Hardware-independent resource projection per homogeneous kernel.
The advection and diffusion columns count logical controlled rotations before
basis decomposition. The last two columns are copy requirements at
$\epsilon_{\rm meas}=0.01$: the histogram value follows
Eq.~\eqref{eq:shot_bound}, whereas the tomography value is the generic
$N/\epsilon_{\rm meas}^{2}$ phase-sensitive proxy~\cite{haah2016sample}. State preparation,
post-selection retries, and the $M=1000$ repetitions are not included in
these per-kernel counts.}
\label{tab:resource_projection}
\begin{tabular}{ccccccc}
\hline
$n_q$ & $N$ & peak qubits & adv.\ rotations & diff.\ rotations &
histogram shots & tomography copies \\
\hline
4  & 16   & 5  & 24   & 32   & $3.75\times10^{4}$ & $1.60\times10^{5}$ \\
6  & 64   & 7  & 96   & 128  & $1.575\times10^{5}$ & $6.40\times10^{5}$ \\
8  & 256  & 9  & 384  & 512  & $6.375\times10^{5}$ & $2.56\times10^{6}$ \\
10 & 1024 & 11 & 1536 & 2048 & $2.5575\times10^{6}$ & $1.024\times10^{7}$ \\
12 & 4096 & 13 & 6144 & 8192 & $1.02375\times10^{7}$ & $4.096\times10^{7}$ \\
\hline
\end{tabular}
\end{table*}
\subsection{Long-horizon integration}
\label{sec:res_longhorizon}

To probe error accumulation and post-selection scaling far beyond the validated horizon, the canonical case was integrated to $T = 50$, i.e.,
$5\times10^{4}$ successive quantum steps. The relative $L_2$ error
against the independently rerun DOP853 reference at $T=50$ is
$1.7606\times10^{-5}$, comparable to its value at $T = 1$,
so no secular error growth is observed in this canonical $T=50$ test across a fifty-fold extension of the horizon. The dynamics over this interval are governed by two well-separated
timescales: the viscous modes relax within $t \lesssim 10$, while the spatial mean, which carries no direct viscous damping at $k = 0$, is damped only
through the shear coupling induced by the varying $U(x)$, giving a slow eigenvalue $\mathrm{Re}\,\lambda = -3.89\times10^{-3}$ ($\tau \simeq 257$)
and a large steady-state response. At $T = 50$ the slow mode has decayed by
the factor $e^{-50/\tau} = 0.82$, and the measured relative distance of the
quantum solution from the steady state, $0.834874$, is consistent with the classical slow-mode trajectory [Fig.~\ref{fig:psucc}(c)]; the energy history
and the $T = 50$ field comparison are shown in Appendix~\ref{app:noise}. The post-selection record remains benign throughout: the cumulative success
probability is $0.022178$, so a no-retry strategy would require approximately
$45.09$ repetitions of the full trajectory, whereas an
ideal per-step RUS accounting gives
$5.000381\times10^{4}$ expected kernel attempts, an
overhead of $0.008\%$. Here, we exclude tomography
and re-preparation. The cumulative $\log_{10}P$ curve
flattens over time as the energy concentrates in weakly dissipated modes
[Fig.~\ref{fig:psucc}(d)], in accordance with
Eq.~\eqref{eq:ancilla_dissipation}.

\flushbottom
\section{Discussion}
\label{sec:discussion}

\subsection{Relation to existing quantum transport solvers}

The results substantiate a specific reallocation of quantum resources relative to prior time-marching solvers. Existing algorithms purchase
generality through embeddings: the Hermitian dilation of
Ref.~\cite{brearley2024quantum} attaches an ancilla register and per-step post-selection to the advective dynamics itself, while Schr\"{o}dinger-embedding approaches~\cite{hu2024quantum,guseynov2026quantum,novikau2025quantum} absorb the entire non-Hermitian operator into a dilated space. The present algorithm instead partitions the dynamics by its physical character. The
skew-symmetric advective generator in Eq.~\eqref{eq:pde}
is implemented unitarily for arbitrary sampled real
$U(x)$, with no ancilla and unit success probability.
The dissipative part, which is genuinely non-unitary, is isolated in the Fourier basis and assigned the minimal quantum resource of one post-selected
ancilla. The affine source update is classical; its
source-corrected output must subsequently be prepared on
the data register. The experiments establish
second-order temporal convergence of the ideal
statevector pipeline and a fixed $n_q+1$-qubit kernel,
not an end-to-end complexity advantage.

\subsection{Post-selection and physical dissipation}

The natural objection to any post-selected primitive concerns its failure rate, and the answer here is quantitatively clear: Eq.~\eqref{eq:ancilla_dissipation} relates the per-step failure probability to the relative viscous dissipation rate to first order in $\Delta t$, with the stated remainder. The measured correlation coefficient differs from unity by less than $4\times10^{-9}$, confirming the predicted temporal relationship. The ideal RUS overhead can therefore be estimated \emph{a priori} from the flow spectrum, before circuit construction. This leads to two observations. First, the cost is partly self-regulating: diffusion preferentially removes the high-wavenumber content responsible for most of the post-selection loss. Consequently, a fifty-fold increase in viscosity reduces the cumulative success probability by less than a factor of three (Table~\ref{tab:nusweep}). Second, under per-step Repeat-Until-Success, a failed ancilla measurement simply triggers re-preparation from the classically stored field; the retry therefore reproduces the same input state without introducing additional algorithmic approximation. The resulting retry overhead remains below $0.14\%$ per step for all viscosities considered and below $0.01\%$ over the $5\times10^{4}$-step horizon. A trade-off identified by the data deserves note: Richardson extrapolation favors larger time steps while the success probability favors smaller $\nu\,\Delta t$, so the two optimizations compete weakly; at the parameters studied, both can be satisfied simultaneously.

\subsection{Cost structure and the hybrid boundary}

Within the audited homogeneous circuit, the advection block accounts for $71\%$--$95\%$ of the CX gates. This does not imply a matching share of end-to-end cost: step-dependent preparation and phase-sensitive tomography were absent from that transpilation. The classical driver reads out and re-prepares the field at every time step; on hardware, the phase-sensitive reconstruction required by the present hybrid feedback boundary cannot be obtained from computational-basis sampling alone and, for generic amplitude reconstruction, requires a number of state copies that grows at least linearly with $N$ at fixed accuracy. Consequently, the end-to-end pipeline does not retain the exponential memory compression of the encoding. No claim of asymptotic advantage is made, consistent with the stated scope of the present work. The measured $138$--$564$ CX figures at $n_q=3$--$4$ belong to a physical-input statevector kernel with two ancillas and no device connectivity. They motivate, but do not by themselves certify, a few-step hardware demonstration.

\subsection{Analysis for end-to-end correctness}
\label{sec:complete_correctness}

The audits above establish continuum consistency, second-order global time
accuracy, fourth-order Richardson behavior, a third-order local splitting
defect, and a quantitative preparation-truncation bound for the ideal
pipeline. A complete correctness statement for a hardware execution must
also connect the normalized quantum state and its stored norm to the physical
field after every classical feedback step. One useful target is the
physical-vector estimate
\begin{align}
\|\mathbf P_hu(T)-\widetilde{\mathbf u}_M\|_2
\leq{}&
C_{\rm stab}(T)\left(C_s h^2+C_tT\Delta t^2\right)
\notag\\
&+C_{\rm stab}(T)\sum_{n=0}^{M-1}\epsilon_{{\rm loc},n}
\notag\\
&+\mathcal E_{\rm noise},
\label{eq:complete_error_budget}
\end{align}
where
$\epsilon_{{\rm loc},n}=\epsilon_{{\rm prep},n}
+\epsilon_{{\rm synth},n}+\epsilon_{{\rm fb},n}$.
Here $\epsilon_{{\rm prep},n}$ is controlled by
Eq.~\eqref{eq:prep_error_bound}, $\epsilon_{{\rm synth},n}$ covers
finite-precision QFT and rotation synthesis, and
$\epsilon_{{\rm fb},n}$ covers phase-sensitive reconstruction, norm
estimation, and re-preparation. The last term is an accumulated bound for
gate, reset, ancilla-readout, and data-readout channels. Conditional on ideal
successful UCRY outcomes, diffusion contributes no additional bias; ideal restart
recovery changes only the expected cost.

The principal missing evidence is therefore a high-probability recursive bound
for $\epsilon_{{\rm fb},n}$, as an amplitude error at one step becomes
the input to the next source update. A refined analysis should use the
skew-Hermiticity of $\mathbf K_{\rm adv}$ and negative semidefiniteness of
$\mathbf D_{\rm diff}$ to replace the conservative
$e^{T\Lambda_h}$ amplification in Eq.~\eqref{eq:global_error_bound} by a
discrete semigroup or logarithmic-norm stability constant. It should then
combine concentration bounds for complex-amplitude reconstruction with
rotation-synthesis and noisy-channel bounds. Finally, the constants should
be tested uniformly under grid refinement, increasing velocity contrast, and
the stiffness parameter $\nu\Delta t\kappa_{\max}^2$. These steps would turn
the separate verified components in Eq.~\eqref{eq:complete_error_budget}
into a full discrete-to-continuum correctness argument. While they are not required
to interpret the present ideal statevector results, they are required
before claiming end-to-end hardware accuracy.

\subsection{Outlook}
The present framework suggests three natural extensions.

First, fully coherent multi-step evolution would eliminate intermediate readout and state re-preparation. This requires a coherent source-injection procedure, for which the sparse Fourier support of the forcing suggests a linear-combination construction over the $K$ active modes, together with coherent norm tracking and observable estimation without full-state tomography, e.g., through amplitude-estimation techniques~\cite{brassard2002quantum}, at the cost of an added ancilla and a second post-selection channel beyond the present diffusion step.

Second, the spatial formulation can be generalized beyond the present one-dimensional periodic setting. The Gray-code construction \cite{savage1997survey} extends to multiple dimensions by applying the edge decomposition independently along each coordinate direction. Non-periodic boundaries can likewise be incorporated through modified boundary stencils, although the exact skew-Hermitian structure is then weakened locally near the boundaries, consistent with the approach of  Bengoechea \emph{et al.}~\cite{bengoechea2026quantum}.

Third, implementation of the complete single-step circuit on quantum hardware would provide a direct test of the predicted post-selection statistics, including the dissipation relation in Eq.~\eqref{eq:ancilla_dissipation}, under realistic device noise. Beyond these extensions, nonlinear transport may be addressed through iterative linearization about the present linear quantum kernel.

\section{Conclusions}
\label{sec:conclusions}

In this work, we presented a hybrid quantum-classical algorithm for the forced skew-symmetric variable-coefficient advection-diffusion equation in Eq.~\eqref{eq:pde}, inspired by common numerical techniques for flow simulations. The homogeneous kernel uses $n_q+1$ logical qubits and was validated against classical integrations of the same semidiscrete model. The construction rests on a single organizing principle that quantum resources are allocated according to the physical character of each operator. Advection is unitary for arbitrary sampled real $U(x)$ because Eq.~\eqref{eq:Kadv} is skew-Hermitian, and is realized as a Gray-code Trotterized circuit with no ancilla and unit success probability. Diffusion, the only genuinely non-unitary element, is applied in the Fourier basis through one post-selected ancilla. The source term is incorporated classically through Strang splitting and succeeds with probability one in every recorded application. An independent operator audit through $N=256$ confirms second-order consistency of the centered skew-symmetric advection stencil.

The numerical evidence supports four conclusions. The ideal statevector pipeline is second-order accurate in time, with observed orders of approximately $2.000$ at both resolutions, and its time-symmetric construction yields an even-power error expansion, so that classical Richardson extrapolation across two circuit kernel trajectories converges at fourth order. The one-step augmented-map defect converges at third order once $\Delta tN^2$ is held fixed under grid refinement, as required to resolve the growing diffusive norm. The first-order ancilla-loss relation in Eq.~\eqref{eq:ancilla_dissipation} exhibits the predicted time dependence in the canonical case, while the fifty-fold viscosity sweep is consistent with the same dissipation-based interpretation. Finally, advection accounts for $71\%$--$95\%$ of the CX count within the transpiled homogeneous-kernel diagnostic, whereas preparation and phase-sensitive reconstruction dominate the unresolved end-to-end boundary.

The principal limitation is the classical readout and re-preparation between steps. Coherent multistep evolution, an explicit tomography experiment, grid convergence of the canonical forced solution against an independent continuum reference, and device-specific dynamic-circuit transpilation are the immediate next tests.

\begin{acknowledgments}
This work was supported by the Air Force Office of Scientific Research
(AFOSR) under Grant No.~FA9550-25-1-0029.
\end{acknowledgments}

\section*{Author Contributions}
M.M.H.\ Akash: Conceptualization, Methodology, Formal analysis,
Writing -- original draft, Writing -- review \& editing.
N.-Q.\ Nguyen: Writing -- review \& editing.
T.\ Dev: Writing -- review \& editing.
Y.\ Chen: Conceptualization, Supervision, Writing -- review \& editing.
H.\ Wu: Conceptualization, Methodology, Supervision, Funding acquisition, Writing -- review \& editing.
K.\ Shoele: Conceptualization, Methodology, Supervision, Funding acquisition, Writing -- review \& editing.

\section*{Data Availability}
The code and result files that support the findings of this article are openly available at \href{https://github.com/akash312010/quantum-linear1d-burgers}{quantum\_linear\_1d\_burgers}.

\section*{Conflict of Interest}
The authors declare no competing interests.
\appendix
\raggedbottom

\section{Robustness and supplementary accuracy audits}
\label{app:robustness}

\subsection{Source complexity and grid resolution}

\begin{figure}[H]
\centering
\includegraphics[width=\linewidth]{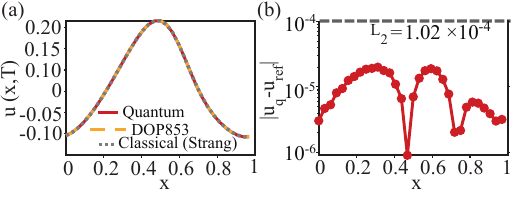}
\caption{Validation at $n_q = 5$: companion to Fig.~\ref{fig:validation}.
(a)~Solution field at $T = 1.0$, $\Delta t = 10^{-3}$, against the DOP853
and matched split-step references. (b)~Pointwise error on a logarithmic scale ($n_q = 5$).}
\label{fig:robustness}
\end{figure}

This subsection verifies that the results of Section~\ref{sec:results} are not
specific to the single-mode forcing or to one grid resolution. Figure~\ref{fig:robustness}(a,b) shows the $n_q = 5$ companion to Figs.~\ref{fig:validation}(a,b); the error structure is identical to the
$n_q = 4$ case, with the magnitudes of Table~\ref{tab:validation}.
The complete validation and convergence studies were repeated with the four-harmonic source $f(x) = 0.5\sum_{j=1}^{4} a_j \sin(2\pi j x/L + \phi_j)$, for which the forcing alone has 8 Fourier modes rather than two. This does not imply that the source-corrected evolving state has $K_n=8$; its support must be audited separately. All temporal signatures reported in the paper are reproduced. The observed convergence orders are $2.001$, $2.000$, and $2.000$ at $n_q = 4$ and $2.000$, $2.000$, and $2.000$ at $n_q = 5$; successive Richardson errors again decrease by factors of $16.0$, falling to $7.8\times10^{-9}$ at $n_q = 4$ and $2.0\times10^{-7}$ at $n_q = 5$ for the finest pair. The error prefactors exceed the single-sine values by only $15\%$ to $25\%$. The source steps succeed with probability exactly one, as they must, and the diffusion post-selection statistics are marginally lower than the single-sine case ($\bar{P}_{\rm succ} = 0.99923$ versus $0.99937$ at $n_q = 5$), consistent with the richer source placing slightly more amplitude in high-wavenumber modes. For the two forcing cases tested, temporal order, extrapolation behavior, and success probability remain quantitatively similar. No general claim about state-preparation cost follows from this test.

\subsection{Richardson extrapolation and accuracy-to-kernel-cost scaling}
\label{app:accuracy_cost}

The fourth-order Richardson behavior established in Sec.~\ref{sec:res_convergence} can be translated into a practical comparison of the number of homogeneous-kernel evaluations required to reach a prescribed accuracy. This supplementary analysis is intended only as an ideal-kernel comparison: state preparation, post-selection retries, measurement, reconstruction, and classical FFT costs remain separate terms in the end-to-end model of Eq.~\eqref{eq:end_to_end_cost}.

\begin{minipage}{\columnwidth}
\centering

\captionof{table}{Accuracy-to-kernel-cost projection based on the fitted convergence laws
$e_{\rm base}=C_2\Delta t^2$ and
$e_{\rm Rich}=C_4\Delta t^4$. Counts are reported over
$T=1$ and include both trajectories required for
Richardson extrapolation.  They represent ideal homogeneous-kernel costs;
state preparation, RUS retries, and readout are accounted for separately in
Eq.~\eqref{eq:end_to_end_cost}.}
\label{tab:accuracy_cost_projection}

\medskip

\begin{tabular}{ccrrr}
\hline
$n_q$ & target $L_2$ & base & Richardson & reduction \\
\hline
4 & $10^{-4}$ & 473   & 108 & $4.38$ \\
4 & $10^{-6}$ & 4725  & 341 & $13.86$ \\
5 & $10^{-4}$ & 1008  & 240 & $4.20$ \\
5 & $10^{-6}$ & 10075 & 756 & $13.33$ \\
\hline
\end{tabular}

\end{minipage}

The base and Richardson errors are fitted as
$e_{\rm base}=C_2\Delta t^2$, $e_{\rm Rich}=C_4\Delta t^4$, using the
convergence data reported in Table~\ref{tab:convergence}. For each target
accuracy, the fitted time step determines the number of time steps
required over $T=1$; a Richardson estimate includes both the fine and
coarse trajectories. Table~\ref{tab:accuracy_cost_projection} reports the
corresponding ideal homogeneous-kernel counts.

The projection shows that the advantage of Richardson
extrapolation grows as the requested accuracy becomes more stringent. At
$n_q=4$, the estimated kernel-count reduction increases from $4.38$ at a
target $L_2$ error of $10^{-4}$ to $13.86$ at $10^{-6}$; the corresponding
factors at $n_q=5$ are $4.20$ and $13.33$. The similar reductions at the
two resolutions are consistent with the observed second- and fourth-order
temporal slopes. These factors quantify only the reduction in ideal
kernel evaluations and are therefore not interpreted as an end-to-end
quantum speedup.

\subsection{Fourier truncation, retained support, and preparation error}
\label{app:prep_audit}

Sparse Fourier preparation introduces a controllable
approximation because coefficients below the threshold $\delta$ are
discarded before loading the state. The relevant quantity is therefore
not the number of Fourier modes in the prescribed source alone, but the
retained support $K_n$ of the complete source-corrected state at each
time step. Because advection and source injection redistribute spectral
amplitude during the evolution, $K_n$ may grow substantially even when
the original forcing is sparse.

\begin{widetableblock}
\captionof{table}{\leftskip0pt\rightskip0pt plus0pt\parfillskip0pt plus1fil\relax
Source-corrected Fourier preparation audit on the $N=64$ run.
$K_n$ is the number of retained modes, $\eta_n$ is the discarded Fourier
norm, and the last column evaluates the normalized one-step bound in
Eq.~\eqref{eq:prep_error_bound}.}
\label{tab:prep_audit}
\medskip
\noindent\makebox[\textwidth][c]{%
\begin{tabular}{cccccc}
\hline
threshold & $\min K_n$ & median $K_n$ & $\max K_n$ &
$\max\eta_n$ & one-step bound \\
\hline
$10^{-4}$  & 11 & 15 & 17 & $1.438\times10^{-4}$ & $2.876\times10^{-4}$ \\
$10^{-6}$  & 15 & 23 & 25 & $2.485\times10^{-6}$ & $4.969\times10^{-6}$ \\
$10^{-8}$  & 29 & 38 & 44 & $3.950\times10^{-8}$ & $7.900\times10^{-8}$ \\
$10^{-10}$ & 42 & 50 & 64 & $2.144\times10^{-10}$ & $4.289\times10^{-10}$ \\
$10^{-12}$ & 52 & 60 & 64 & $1.958\times10^{-12}$ & $3.915\times10^{-12}$ \\
\hline
\end{tabular}%
}
\end{widetableblock}

Table~\ref{tab:prep_audit} shows the expected tradeoff between
truncation error and preparation complexity. At the relatively loose
threshold $10^{-4}$, the $N=64$ trajectory retains between $11$ and $17$
modes, with median $K_n=15$. Tightening the threshold progressively
enlarges the support: at $10^{-10}$, the maximum support reaches all $64$
Fourier modes, and at $10^{-12}$ the median support is already $60$. At
the same time, the discarded norm and the corresponding one-step bound
decrease from $\mathcal O(10^{-4})$ to $\mathcal O(10^{-12})$.

This audit makes explicit the principal tradeoff of
the sparse preparation strategy. Aggressive truncation can substantially
reduce the logical preparation count $\mathcal O(K_n n_q)$, but
sufficiently stringent accuracy requirements can drive $K_n$ toward $N$,
eliminating the sparsity advantage. For this reason, the end-to-end
resource estimates retain $K_n$ explicitly rather than assuming that the
evolving state inherits the sparsity of the forcing.

\section{Measurement statistics, sampling-noise model, and long-horizon supplements}
\label{app:noise}

\subsection{Sampling-circuit noise model}
\label{app:sampling_noise}

\begin{figure}[H]
\centering
\includegraphics[width=\columnwidth]{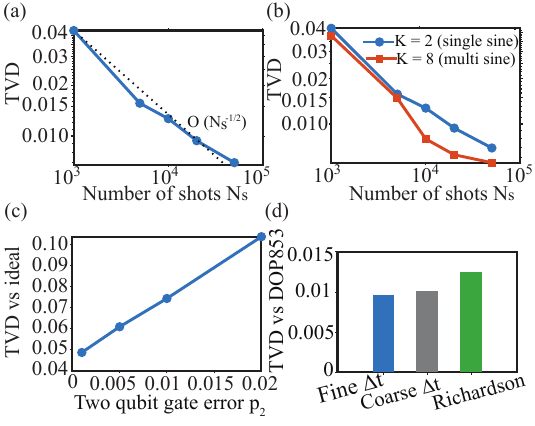}
\captionsetup{
    width=\linewidth,
    singlelinecheck=false
}
\caption{
\protect\justifying
Finite-shot and sampling-noise evaluation at $n_q = 4$. (a)~Total variation
distance (TVD) between sampled and exact final-state distributions, following the
$N_s^{-1/2}$ reference scaling. (b)~The same measurement for the two-mode ($K = 2$) and four-harmonic ($K = 8$) sources; no additional sampling
penalty is observed for the tested four-harmonic source. (c)~Distributional error under the depolarizing-plus-readout noise model at $2\times10^{4}$ shots as a function of the two-qubit error rate. (d)~Probability-space Richardson extrapolation at $2\times10^{4}$ shots,
illustrating the shot-noise amplification below the extrapolation threshold.}
\label{fig:shots}
\end{figure}

This experiment characterizes the sensitivity of the reconstructed
probability distribution to finite-shot and sampling-circuit noise; it does
not constitute a device-level noisy execution of the complete D-A-D quantum
kernel. The noise model combines depolarizing errors of
probability $p_1 = 10^{-3}$ on single-qubit gates and $p_2$ on CX gates with
a symmetric readout error of $2\%$ per qubit, applied to the transpiled
sampling circuit at $2\times10^{4}$ shots. Sweeping
$p_2 \in \{0.001, 0.005, 0.01, 0.02\}$ moves the TVD only from
$4.9\times10^{-2}$ to $1.0\times10^{-1}$ [Fig.~\ref{fig:shots}(c)], a factor
of two across a twenty-fold change in gate error. No separate
readout-only versus gate-only control was performed, so this sweep does not
by itself isolate the dominant channel; the relatively weak variation across
the tested two-qubit depolarizing-error range, in the presence of a fixed
$2\%$ readout error, suggests that readout error contributes an important
error floor in this sampling experiment.

\subsection{Ancilla--dissipation remainder refinement}
\label{app:dissipation_remainder}

The
magnitude of the ancilla--dissipation identity (Eq.~\eqref{eq:ancilla_dissipation}) is tested separately by refining $\Delta t$ and tracking the
maximum remainder of the step-centered evaluation, $r_n=(1-P_n)-\Delta t\,\bar\varepsilon_n/\bar E_n$, reported
in Table~\ref{tab:postselection_remainder}.

Table~\ref{tab:postselection_remainder} tests the magnitude, rather than only
the correlation, of Eq.~\eqref{eq:ancilla_dissipation}. The median order is
$2.00006$, and $\max_n|r_n|/\Delta t^2$ approaches $0.6103$. This verifies
the stated quadratic remainder over a factor-eight refinement in time.

\begin{table}[H]
\centering
\caption{Refinement of the maximum ancilla--dissipation remainder of the
step-centered evaluation,
$r_n=(1-P_n)-\Delta t\,\bar\varepsilon_n/\bar E_n$, with
$\bar E_n=\tfrac12(E_n+E_{n+1})$ and
$\bar\varepsilon_n=\tfrac12(\varepsilon_n+\varepsilon_{n+1})$. The observed order is
computed between successive time steps.}
\label{tab:postselection_remainder}
\begin{tabular}{ccc}
\hline
$\Delta t$ & $\max_n|r_n|$ & order \\
\hline
$0.0040$ & $9.7655\times10^{-6}$ & -- \\
$0.0020$ & $2.4413\times10^{-6}$ & $2.00004$ \\
$0.0010$ & $6.1029\times10^{-7}$ & $2.00009$ \\
$0.0005$ & $1.5257\times10^{-7}$ & $2.00006$ \\
\hline
\end{tabular}
\end{table}

\subsection{Richardson extrapolation under finite-shot sampling}
\label{app:shot_richardson}

Applying
$(4p_{\Delta t} - p_{2\Delta t})/3$ to \emph{measured} probability
distributions at $2\times10^{4}$ shots yields a TVD of $1.25\times10^{-2}$
against the reference \cite{temme2017error,kandala2019error}. This is worse than the $9.6\times10^{-3}$ of the fine-step
measurement alone [Fig.~\ref{fig:shots}(d)]. This trend is expected since
the linear combination amplifies independent shot noise by
$\sqrt{16+1}/3 \approx 1.37$, and indeed
$1.37 \times 9.6\times10^{-3} \approx 1.32\times10^{-2}$ reproduces the
measured value. Probability-space extrapolation is therefore beneficial only
when the sampling error lies below the splitting-error difference between the two time steps; at the present parameters, a shot-noise estimate suggests this would require shot counts roughly two
orders of magnitude beyond $2\times10^{4}$. The statevector-level extrapolation of Section~\ref{sec:res_convergence} is unaffected by this
threshold.

\subsection{Long-horizon supplementary diagnostics}
\label{app:longhorizon_supp}

\begin{figure}[t]
\centering
\includegraphics[width=\linewidth]{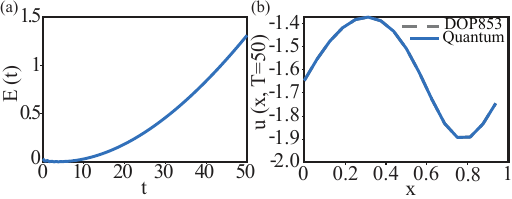}
\captionsetup{
    width=\linewidth,
    singlelinecheck=false
}
\caption{\protect\justifying Long-horizon supplements to Sec.~\ref{sec:res_longhorizon}.
(a)~Energy history to $T = 50$ as the weakly damped quasi-mean mode charges
toward the forced steady state. (b)~Quantum and DOP853 fields at $T = 50$
($5\times10^{4}$ steps), agreeing to a relative $L_2$ error of
$1.8\times10^{-5}$.}
\label{fig:longhorizon_supp}
\end{figure}
Figure~\ref{fig:longhorizon_supp} completes
the $T = 50$ study of Section~\ref{sec:res_longhorizon}. The energy grows
monotonically after the initial viscous transient as the quasi-mean mode,
which carries no direct viscous damping, charges toward its large
steady-state amplitude through the shear coupling
$d\langle u\rangle/dt = \tfrac12\langle u\,\partial_x U\rangle$; the system matrix is
well conditioned (condition number $5.2\times10^{3}$), with slowest
eigenvalue $\mathrm{Re}\,\lambda = -3.89\times10^{-3}$ setting the
$\tau \simeq 257$ approach timescale. The quantum and DOP853 fields at
$T = 50$ agree to a relative $L_2$ error of $1.8\times10^{-5}$ after
$5\times10^{4}$ quantum steps.
\section{Gray-code labeling and even--odd Trotter construction}
\label{app:graycode}

The Gray-code relabeling used in Sec.~\ref{sec:dad} maps
nearest-neighbor grid couplings onto transitions between
computational-basis states differing in only one qubit. In ordinary
binary ordering this property is not guaranteed; for example, the $n_q=2$
transition $m=1\to2$ corresponds to $01\to10$, which changes two bits.

\begin{table}[H]
\centering
\caption{Binary and Gray-code labeling of grid points for
$n_q=2$ ($N=4$). Consecutive Gray-code labels differ in exactly
one bit, whereas the binary transition $m=1\rightarrow2$ changes
two bits simultaneously.}
\label{tab:gray_example}
\begin{tabular}{ccc}
\hline
Grid index $m$ & Binary & Gray code $g(m)$ \\
\hline
0 & 00 & 00 \\
1 & 01 & 01 \\
2 & 10 & 11 \\
3 & 11 & 10 \\
\hline
\end{tabular}
\end{table}

Gray-code relabeling determines how an individual nearest-neighbor
edge is encoded as a controlled single-target rotation
(Fig.~\ref{fig:advection}). The even-odd partition used in
Eq.~\eqref{eq:adv_trotter}, by contrast, determines the symmetric Trotter
ordering of those edge operations. Consequently, the two constructions serve
distinct purposes: Gray code simplifies edge encoding, whereas the
even-odd-even sequence supplies second-order temporal accuracy.

\onecolumngrid
\section{Detailed circuit-resource audit}
\label{app:resource_audit}
Table~\ref{tab:transpilation_audit} gives the raw transpilation counts underlying Fig.~\ref{fig:cost}. The diagnostic circuit uses two diffusion ancillas and a leading QFT, whereas the production Fourier-input construction reuses one ancilla and omits that QFT; these counts therefore characterize the homogeneous D-A-D kernel in isolation, not an end-to-end hardware cost — step-dependent state preparation, classical source arithmetic, phase-sensitive reconstruction, and device connectivity/routing are not included.
\small
\setlength{\tabcolsep}{4.5pt}
\captionof{table}{\leftskip0pt\rightskip0pt plus0pt\parfillskip0pt plus1fil\relax
Transpilation audit of the physical-input homogeneous D-A-D
statevector circuit (basis
$\{\mathrm{CX}, \sqrt{X}, R_z, X\}$, optimization level 3). This audit
uses two diffusion ancillas and includes a leading QFT; the production
Fourier-input construction can reuse one ancilla and omits that QFT.
The preparation column is the circuit for $f$ alone and is not the
preparation cost of the generally dense source-corrected state. The UCRY and
QFT columns are counts for one block and are not additive components of the
reported total.}
\label{tab:transpilation_audit}
\medskip
\noindent\makebox[\textwidth][c]{%
\begin{tabular}{ccccccccc}
\hline
$n_q$ & sim.\ qubits & depth & CX total & prep($f$) & UCRY & QFT & advection & adv.\ share \\
\hline
3 & 5 & $454$  & $138$  & $1$ & $8$  & $6$  & $98$   & $71\%$ \\
4 & 6 & $1809$ & $564$  & $2$ & $16$ & $12$ & $484$  & $86\%$ \\
5 & 7 & $2450$ & $1297$ & $3$ & $30$ & $20$ & $1157$ & $89\%$ \\
6 & 8 & $6788$ & $4072$ & $4$ & $52$ & $30$ & $3848$ & $95\%$ \\
\hline
\end{tabular}%
}
\vspace{0.3em}

\twocolumngrid

\end{document}